\documentclass[10pt,journal]{IEEEtran}
\ifCLASSINFOpdf
\else
\fi
\usepackage{optidef}
\usepackage{upgreek}
\usepackage{graphicx} 
\usepackage{mdframed}

\usepackage{lettrine}
\usepackage{algorithm}
\usepackage{tikz}
\usetikzlibrary{positioning,arrows.meta}
\usepackage{mathrsfs}
\usepackage{caption}
\usepackage[noadjust]{cite}
\usepackage{etoolbox}
\apptocmd{\equation}{\vspace{-0.3em}}{}{}
\pretocmd{\endequation}{\vspace{-0.2em}}{}{}
\usepackage{comment}
\usepackage{algpseudocode}
\usepackage[english]{babel}
\DeclareMathOperator*{\argmax}{arg\,max}

\usepackage{mathtools}
\usepackage{soul}
\usepackage{color}

\newtheorem{theorem}{Theorem}[section]

\newtheorem{proposition}{Proposition}
\newtheorem{corollary}[theorem]{Corollary}

\newenvironment{proof2}[1][Proof]{\par\noindent\textit{#1.}\ }{\hfill$\square$\par}

\usepackage{tabularx,booktabs}
\usepackage{xcolor}
\usepackage[utf8]{inputenc}
\usepackage{amssymb}
\usepackage{amsmath}
\usepackage{graphicx,amsfonts,booktabs,multirow,array,comment}
\usepackage{color, colortbl}
\definecolor{Gray}{gray}{0.9}
\definecolor{LightCyan}{rgb}{0.88,1,1}
\definecolor{newgreen}{rgb}{0.0,0.55,0.0}
\usepackage[normalem]{ulem}

\usepackage{titlesec}

\titlespacing*{\section}
  {0pt}{1.0ex plus 0.2ex minus 0.2ex}{0.5ex}

\titlespacing*{\subsection}
  {0pt}{1.5ex plus 0.3ex minus 0.2ex}{0.6ex}

\titlespacing*{\subsubsection}
  {0pt}{0.6ex plus 0.2ex minus 0.2ex}{0.3ex}
\title{%
Doppler-Aware Meta-Learning for Evolutive STAR-RIS in Cognitive Autonomous Networks}

\author{Noha~Hassan,~\IEEEmembership{SMIEEE,}
Maysa~Yaseen,~\IEEEmembership{MIEEE,}
        Xavier~Fernando,~\IEEEmembership{SMIEEE,}
        and~Halim~Yanikomeroglu,~\IEEEmembership{FIEEE}%
        \vspace{-0.5em}
\thanks{This work has been submitted to the IEEE JSTSP for possible publication. Copyright may be transferred without notice, after which this version may no longer be accessible}
\thanks{N. Hassan and H. Yanikomeroglu are with the Department of Systems and Computer Engineering,
Carleton University, Ottawa, ON, Canada. N. Hassan is also with the Department of Electrical and Computer Engineering,
Dalhousie University, NS, Canada.
(e-mail: nohahassan@sce.carleton.ca.ca; noha.hassan@dal.ca; halim@sce.carleton.ca).}
\thanks{M. Yaseen is with the Department of Electrical and Computer Engineering,
Lakehead University, Barrie, ON, Canada
(e-mail: myaseen@lakeheadu.ca ).}
\thanks{X. Fernando is with the Department of Electrical, Computer, and Biomedical Engineering,
Toronto Metropolitan University, Toronto, ON, Canada and also with the Karunya Institute of Technology and Sciences
(e-mail: fernando@torontomu.ca).}}
\begin{document}

\maketitle
\markboth{IEEE JSTSP Special Issue on Autonomous and Evolutive Optimization in Networked AI}%
{Hassan \MakeLowercase{\textit{et al.}}: }
\pagestyle{headings}

\begin{abstract}
In real-time optimization of simultaneous transmission and reflection reconfigurable intelligent surfaces (STAR-RIS), limitations arise due to Doppler effects and reduced adaptation speed. Existing RIS memoryless surface models use quasi-static optimization and incur high latency, which limits their application in cognitive autonomous networks.
To overcome this limitation, we propose a stateful RIS model that includes an element-wise memory, which locally stores and updates its phase history across coherence intervals. Hence, the metasurface becomes a stateful surface with element-level memory that enables the exploitation of temporal channel correlations. 
This element-wise memory is complemented with a meta-parameter, yielding faster convergence and giving rise to hierarchical adaptation in gradient and episodic timescales.
The proposed framework supports autonomous and experience-driven learning, which enables the STAR-RIS architecture and memory to adapt according to the Doppler effect, as seen in cognitive communication systems.
Results demonstrate an improvement in spectral efficiency and adaptation speed across various mobility patterns, fading models, interference levels, and array sizes. An analytical field-programmable gate array (FPGA) latency estimate indicates that the critical path fits within the coherence window.
\end{abstract}


\IEEEpeerreviewmaketitle
\vspace{-1em}
\section{Introduction}
\lettrine{S}{imultaneously} transmitting and reflecting reconfigurable intelligent surfaces (STAR-RISs) have the potential to enhance coverage, spectral efficiency, and energy consumption by providing higher degrees of freedom, thus becoming appealing in densely populated urban/indoor settings where the line-of-sight (LoS) paths are restricted \cite{cao2026stacked}.
Optimizing the element phases provides constructive/destructive interference patterns where the collective phase configuration creates directional beams toward specific directions. In practical deployments, the wireless environment is highly dynamic with mobility, multipath fading, and uncertain channel state information (CSI). Hence, the phase optimization problem is highly non-convex and NP-hard due to the discrete nature of phase shifts, mutual coupling effects, and hardware constraints \cite{yang2026analysis, wang2026performance, ahmed2023survey}. Solving the optimization problem in real time is a prerequisite for cognitive autonomous networks, where the infrastructure must sense, learn, and reconfigure itself continuously.

Classical optimization algorithms suffer from high computational complexity, slow convergence rate, lack of flexibility, and non-interpretability \cite{oh2026unified, papazafeiropoulos2023cooperative, yao2026deep}. 
In this work, we propose \emph{Meta2L}, a hierarchical episodic meta-learning framework for STAR-RIS optimization for dynamic wireless environments. Our episodic learning framework includes two layers; an intra-episode gradient-based RIS phase optimization layer and a cross-episode meta-policy learning layer of adaptation strategies and experiences. Conventional algorithms assume a rapid jump in phase configurations across various coherence interval periods. To model the smooth transition, we represent a stateful model in which the phase configuration at each coherence interval preserves part of its previous state, where temporal inertia and element resistance contribute to this state-keeping effect.
Another layer of memory is maintained for each element, thereby allowing individual elements to have their own phase memory. This converts the surface from being a passive programmable reflector to a stateful surface that possesses element memory.
The element memory is a logical state, resembling a register in the driver of a unit cell or an entry per element in a controller table. This is updated on an episode-by-episode basis according to the timescale of channel coherence, as opposed to the timescale of individual gradient updates, allowing for a clean distinction between intra-episode learning and inter-episode memory.

The feasibility of our approach is illustrated by two hardware-aware evaluations. In the first one, we conduct a latency estimate where each of the operations performed along the critical path for the Meta2L algorithm is mapped analytically by cycle counting to an existing field-programmable gate array (FPGA) chip to check whether the full optimization cycle fits within the channel coherence window. In the second one, we introduce varactor circuit impairments by applying phase quantization. Each of the continuous phases is mapped to the nearest hardware-representable discrete value and a settling delay that captures the transient dynamics of a practical phase shifter as a weighted average between the past and desired phases.
We also introduce an online mirror-descent-based tag-weight learning algorithm that identifies which contextual labels are most predictive of relevant memory retrieval. The update of the tag weights via mirror descent helps warm-start initialization as it increases the probability of fetching appropriate past configurations. As a result, fewer iterations of the gradient descent algorithm are needed, and the causal filter avoids using inappropriate warm-start initializations.

The summary of our key contributions includes the development of: 

(i) A stateful surface-inertia model in which the phase configuration evolves according to learned temporal dynamics. This creates a direct connection with the Gauss–Markov channel model and allows the system to predict phase changes, reducing the need for frequent control updates.

(ii) An element-level memory model in which each element keeps a record of its previous phase states at the episodic timescale. This gives the surface a form of state-based computation rather than treating it simply as a passive reflector.

(iii) A stationary point formulation for phase optimization via Lagrangian relaxation based on signal-to-interference-plus-noise ratio (SINR) gradients. 

(iv) An online mirror-descent tag-weight learning algorithm over the probability simplex. 

(v) A Doppler-aware warm-start algorithm that exploits the Gauss--Markov channel model for phase initialization at each new episode from a closed-form Doppler-weighted interpolation of the two most relevant past episodes. 

(vi) A hardware-aware evaluation, consisting of an analytical FPGA latency estimate, which indicates that the proposed algorithm meets real-time deployment constraints, and a simulation showing that stateful control overtakes memoryless RIS under realistic phase quantization and varactor settling delay. 

These contributions realize an evolutive, experience-driven learning surface whose phase configuration, memory states, and meta-parameters evolve continuously by episodic learning, in line with the autonomy and adaptability principles of networked artificial intelligence (AI) for cognitive communications.

The remainder of this paper is organized as follows: Section~\ref{sec1} presents the related work; Section~\ref{sec2} covers the system model; Section~\ref{sec3} presents the proposed algorithm and the problem formulation, including the new theoretical contributions; Section~\ref{sec5} covers the results; and Section~\ref{sec6} concludes the paper.
\vspace{-0.8em}
\section{Related Work}
\label{sec1}
STAR-RIS technology stands out as a key enabler capable of intelligently reconfiguring electromagnetic propagation by optimizing phase shifts to extend coverage, improve spectral efficiency, and reduce energy consumption. Rapidly changing channel conditions, user demands, and interference patterns in dynamic STAR-RIS environments expose the limitations of classical phase-shift optimization, particularly in terms of convergence speed, scalability, and adaptability to non-stationary channels.

The authors in \cite{du2022performance} propose a two-phase channel estimation scheme for RIS-assisted terahertz systems with beam split. The phases are then designed for the current interval only, without reusing the configurations found in earlier intervals.
The work in \cite{praia2021phase} realizes a liquid-crystal RIS for wide-angle terahertz beam steering, whose elements cannot jump instantly to a new phase.
The authors in \cite{zappone2020optimal} configure the RIS with a modified Aquila optimizer, a swarm intelligence optimization (SIO) algorithm, which needs many iterations and careful tuning and does not exploit gradients.
Meta-learning algorithms, particularly model-agnostic meta-learning (MAML) \cite{ chen2025maml} and its variants, have been shown to enable faster adaptation and improved handling of non-convex optimization landscapes in learning-based systems. Nevertheless, they are limited to static problems and require task-specific training data for each new environment. Prior literature on bi-level meta-learning methods does not address regret analysis for meta-policy learning, improvement in convergence rate through exploitation of temporal channel structure, or mechanisms for ensuring robustness against erroneous model predictions. 
STAR-RIS beamforming under CSI uncertainty using alternating optimization (AO) is addressed in \cite{11313224}, which separates beamforming and RIS phases, limiting scalability. The authors in \cite{yang2022meta} jointly learn probing-beam patterns and predict beams for mmWave massive MIMO with neural networks, but the model is scenario-specific and carries no state across intervals.
RIS beamforming using deep reinforcement learning (DRL) has been explored \cite{zhong2022hybrid, yao2026deep, iqbal2026soft}; yet, it is faced with sample complexity issues and poor convergence due to mobility-induced non-stationarity. 
Online mirror descent over the probability simplex \cite{lan2023policy} provides optimal regret guarantees for weight learning; however, it has not yet been applied to tag-based similarity retrieval in episodic meta-learning for wireless systems.
Meta-learning has also been applied to STAR-RIS design, where Javadi et al. \cite{javadi2025meta} use meta-learning to allocate resources in an uplink NOMA system aided by multiple active STAR-RISs. Eghbali et al. \cite{eghbali2024beamforming} design the beamforming of a STAR-RIS-aided integrated sensing and communication system with meta deep reinforcement learning. Amiri \emph{et al.} \cite{amiri2025resource} apply meta-learning to resource allocation in a STAR-RIS-aided system with simultaneous wireless information and power transfer and rate-splitting multiple access. These works use meta-learning across tasks, such as different user configurations or channel realizations, to speed up adaptation to a new task. In both cases, however, the task is treated as quasi-static. Nevertheless, they do not carry state from one coherence interval to the next, use a Doppler-dependent warm start, or account for surface inertia, which are the main differences considered in this paper. 
Applying these methods directly to the phase-tracking problem considered here would therefore change their original problem formulation and reduce them to a gradient-based meta-learning method or a deep reinforcement learning policy. These correspond to the GMLB and DRL-PPO baselines used in Section~\ref{sec5}. For this reason, we compare against these two baseline families, as well as the simple phase-reuse strategies described in Section~\ref{sec5}, while discussing the two related works here.
In addition, proximal policy optimization (PPO) \cite{Schulman2017ProximalPO} and soft actor-critic (SAC) \cite{iqbal2026soft} policies need to interact with the environment thousands of times for each episode while not being able to transfer any information between episodes. 
Deep reinforcement learning with proximal policy optimization (DRL-PPO) \cite{kamal2025optimizing} can deal with time-varying channels but also can not transfer knowledge among episodes.
In \cite{dai2023two}, the authors design the phase shifts of RISs only based on slowly-varying statistical CSI, meaning the phase shifts do not need to be redesigned for a long time, but only when the statistical CSI changes. This decreases power consumption and feedback overhead compared to the instantaneous CSI-based scheme that redesigns phases in each coherence interval.
Finally, the authors in \cite{zardi2024memory} propose an evolutionary method for dynamic RIS control that implements memory-enhanced search mechanisms to exploit the time/space correlation of communication environments. However, their memory is centralized, not an element-level state. They also use evolutionary search rather than gradient-based meta-learning, and they do not model temporal inertia.
The key contribution of the networked AI approach used in this paper is its filling of the aforementioned gap by treating the optimization of RIS as an experience-based process driven by the system’s own accumulated experience. This is the central motivation of the networked AI perspective adopted in this work.
\vspace{-0.4em}
\section{System Model}
\label{sec2}
An outdoor wireless uplink multi-user system consisting of $M$ access 
points (APs), each equipped with $Q$ receiver antennas configured 
as a uniform linear array (ULA), serving $K$ single-antenna 
user equipments (UEs) is considered in our model.
A context-aware adaptive STAR-RIS with $N$ reconfigurable elements is 
mounted on a building facade, where UEs on the same side as the APs are 
served via reflection and UEs on the opposite side are served via transmission. All channels are modeled using a time-varying Rician fading model to capture both line-of-sight and scattered components, as well as the effects of UE mobility and Doppler spread.
\vspace{-0.7em}
\subsection{Channel Model}
Each UE $k$ moves with velocity $v_k$ in the direction $\varphi_k$ with 
respect to the AP boresight; consequently, the Doppler shift $f_{D,k}$ is defined as $f_{D,k}=v_k\cos(\varphi_k)/\lambda$. Over consecutive channel coherence intervals of duration 
$T_c = 1/(2 f_{D,\max})$, with $f_{D,\max} = \max_k f_{D,k}$, the AP-RIS channel coefficients evolve according to the first-order Gauss--Markov model {\cite{baddour2005autoregressive}} as follows:
\begin{equation}
\mathbf{H}_{{q}}(t) = \rho_c\, \mathbf{H}_{{q}}(t-1) + 
\sqrt{1-\rho_c^2}\, \boldsymbol{\mathcal{E}}_{{q}}(t),
\label{eq:gauss_markov}
\end{equation}
where $\rho_c = J_0(2\pi f_{D,\max} T_s)$ is the temporal correlation
coefficient, $J_0(\cdot)$ is the zeroth-order Bessel function of the
first kind, $T_s$ is the interval between two consecutive channel updates (episodes), $\mathbf{H}_{{q}}(t) \in \mathbb{C}^{Q \times N}$ denotes the uplink
channel matrix from the $N$ elements to AP $q$'s $Q$ receive antennas at
coherence interval $t${, whose $n$-th column $[\mathbf{H}_q(t)]_{:,n}$ is the channel from element $n$}, and $\boldsymbol{\mathcal{E}}_{{q}}(t) \sim \mathcal{CN}(\mathbf{0}, 
\mathbf{I}_{Q\times N})$ is the noise matrix which is 
independent across time slots. Let $\mathbf{h}_{{k}}^{\mathsf{r}}(t) \in \mathbb{C}^{N \times 1}$ and 
$\mathbf{h}_{{k}}^{\mathsf{t}}(t) \in \mathbb{C}^{N \times 1}$ denote the uplink 
channel vectors from UE $k$ to the STAR-RIS for the reflection ($\mathsf{r}$) and 
transmission ($\mathsf{t}$) paths, respectively, at time $t$. It is worth noting that the same model applies for the UE-to-RIS 
channels $\mathbf{h}_{{k}}^{\mathsf{r}}(t)$ and $\mathbf{h}_{{k}}^{\mathsf{t}}(t)$ with 
their respective Doppler-derived correlation coefficients.
The Gauss--Markov assumption is crucial for temporal warm starting as it enables the meta-policy to start its search process 
from a point that is {close to a stationary point, as stated in  Section~\ref{sec3}}, thereby ensuring fewer
gradient steps to find an $\epsilon$-stationary point, where $\epsilon$ is an optimality tolerance.
Due to variations in $\rho_c$ caused by changes in UE velocity, an optimal phase configuration at time $t$ may become suboptimal within a single coherence period for high-mobility UEs. This highlights the need for rapid adaptation, which is enabled by the proposed framework.

The reflection and transmission phase shift 
vectors of the STAR-RIS are 
$\boldsymbol{\phi}^\mathsf{r}(t)$ and 
$\boldsymbol{\phi}^\mathsf{t}(t) \in \mathbb{R}^N$ 
respectively, with 
$\phi_n^{\mathsf{r}}(t), \phi_n^{\mathsf{t}}(t) 
\in [0, 2\pi)$ for all $n$.
The concatenated phase vector is expressed as 
\begin{equation}
\boldsymbol{\phi}(t) \triangleq 
[(\boldsymbol{\phi}^{\mathsf{r}}(t))^T, (\boldsymbol{\phi}^{\mathsf{t}}(t))^T]^T 
\in \mathbb{R}^{2N},
\label{eq:phi_def}
\end{equation}
whose temporal evolution is governed by the stateful model in~\eqref{eq:stateful_ris}.
The instantaneous realizations are expressed in \eqref{eq:H_qn} 
and \eqref{eq:h_nk}, 
\begin{figure*}[!t]
\centering
\begin{equation}
\mathbf{H}_{{q}}(t) = \sqrt{\zeta_{{q}}} \!\left(
\sqrt{\frac{\kappa_{{q}}}{1 + \kappa_{{q}}}} \mathbf{H}_{{q}}^{\mathrm{LoS}}(t) +
\sqrt{\frac{1}{1 + \kappa_{{q}}}} \mathbf{H}_{{q}}^{\mathrm{NLoS}}(t)  
\right) \in \mathbb{C}^{Q \times N}
\label{eq:H_qn}
\end{equation}
\vspace{0.3em}
\begin{equation}
\mathbf{h}_{{k}}^{\mathsf{r}/\mathsf{t}}(t) = \sqrt{\zeta_{{k}}^{\mathsf{r}/\mathsf{t}}} \!\left(
\sqrt{\frac{\kappa_{{k}}^{\mathsf{r}/\mathsf{t}}}{1 + \kappa_{{k}}^{\mathsf{r}/\mathsf{t}}}}  
\mathbf{h}_{{k}}^{\mathrm{LoS},\mathsf{r}/\mathsf{t}}(t) +
\sqrt{\frac{1}{1 + \kappa_{{k}}^{\mathsf{r}/\mathsf{t}}}}  
\mathbf{h}_{{k}}^{\mathrm{NLoS},\mathsf{r}/\mathsf{t}}(t) \right) \in \mathbb{C}^{N \times 1}
\label{eq:h_nk}
\end{equation}
\vspace{0.1em}
\hrule
\vspace{-1.5em}
\end{figure*}
where $\zeta_{{q}}$ and $\zeta_{{k}}^{\mathsf{r}/\mathsf{t}}$ are large-scale path
loss coefficients (slowly varying), $\kappa_{{q}}$ and $\kappa_{{k}}^{\mathsf{r}/\mathsf{t}}$ are
Rician $K$-factors,  
and $\mathbf{H}_{{q}}^{\mathrm{NLoS}}(t)$ 
and $\mathbf{h}_{{k}}^{\mathrm{NLoS},\mathsf{r}/\mathsf{t}}(t)$  are the time-varying NLoS 
components. 
The LoS components follow standard ULA 
geometry, where 
$\mathbf{H}_{{q}}^{\mathrm{LoS}}(t)$ and 
$\mathbf{h}_{{k}}^{\mathrm{LoS},\mathsf{r}/\mathsf{t}}(t)$ 
are formed from array steering vectors 
$\mathbf{a}_N(\cdot)$ evaluated at the 
corresponding angles of arrival and departure.
The STAR-RIS reflection and transmission matrices follow the energy-splitting model \cite{mu2021simultaneously}, in which element $n$ divides the incident power between reflection and transmission with ratio $\eta_n\in[0,1]$, and are expressed as
\begin{equation}
\mathbf{\Psi}^\mathsf{r}(t) \triangleq \operatorname{diag}\!\left({\sqrt{\eta_n}}\,e^{j\phi_n^\mathsf{r}(t)}\right){_{n=1}^{N}} \in \mathbb{C}^{N\times N}, \label{eq:Psi_r}
\end{equation}
\begin{equation}
\mathbf{\Psi}^\mathsf{t}(t) \triangleq \operatorname{diag}\!\left({\sqrt{1-\eta_n}}\,e^{j\phi_n^{\mathsf{t}}(t)}\right){_{n=1}^{N}} \in \mathbb{C}^{N\times N}. \label{eq:Psi_t}
\end{equation}
It then results to $|[\mathbf{\Psi}^\mathsf{r}]{nn}|^2+|[\mathbf{\Psi}^\mathsf{t}]{nn}|^2=1$ for a passive element. This means that the reflected and transmitted signals share the total power instead of each having its own unit-gain path. The ratios $\eta_n$ are fixed by the deployment and are not optimized in this work. We set $\eta_n=1/2$ for all $n$ in the simulations. This scales the effective channels of all users by the same factor $1/\sqrt{2}$, thus the optimal phases are unchanged. The sum-rates reported in Section~\ref{sec5} therefore correspond to an energy-splitting surface with $\eta_n=1/2$, with the users transmitting 3~dB more power than would be needed with unit amplitude on each path. For lossless passive elements, the reflection and transmission phases can also be coupled \cite{zhong2022hybrid}. We use independent phases here, corresponding to elements with separate tuning circuits for the two responses. If the phases are coupled, we can write $\phi_n^{\mathsf{t}}=\phi_n^{\mathsf{r}}+\Delta_n$ with a fixed offset $\Delta_n$. This cuts down the control variables by half, without altering the memory mechanism or warm start approach.
Each element $n$ operates in one of three adaptive modes that are defined 
according to a per-element gain coefficient $\alpha_n(t)$. The first is a unit-gain mode, where the unit-gain phase control ($\alpha_n(t) = 1$). In this case, the dominant power cost arises from the PIN/varactor diode biasing circuit. 
The second is the semi-active mode, with a phase control of low-gain amplification ($\alpha_n(t) = \alpha_{\mathrm{act}}$), where $\alpha_{\mathrm{act}} > 1$ represents the fixed amplification gain, which is determined by the hardware specification 
of the semi-active element. This is powered by 
an external energy source, incurring a higher per-element power cost. The third is a sleep mode, where the element is temporarily deactivated ($\alpha_n(t) = 0$). A deactivated element contributes zero to the beamformed output, consistent with a shorted or absorbing element state.
The semi-active mode follows the reflection-type amplifiers of active RIS \cite{long2021active,zhang2023active}. It is subject to two physical constraints that a passive element does not have. 
First, the extra output power is taken from the external supply, and the amplified power is constrained by the budget $\sum_{n} \alpha_n^2(t) \, \mathbb{E}|x_n^{\mathrm{in}}(t)|^2 \le P_{\mathrm{A}}$, where $x_n^{\mathrm{in}}(t)$ is the signal incident to the element $n$ and its cost is accounted for via $P_{\mathrm{act}}$.
Second, the amplifier adds thermal noise $\mathbf{n}_{\mathrm{s}}(t)\sim\mathcal{CN}(\mathbf{0},\sigma_{\mathrm{s}}^2\mathbf{I}_N)$, which reaches AP $q$ as $\mathbf{H}_q(t)\mathbf{A}(t)\mathbf{n}_{\mathrm{s}}(t)$ and adds $\sigma_{\mathrm{s}}^2\|\mathbf{H}_q(t)\mathbf{A}(t)\|_F^2/Q$ per antenna to the noise power in (\ref{eq:SINR}). A modest gain $\alpha_{\mathrm{act}}=2$ keeps both terms small. The numerical outcomes for all results presented in Section~\ref{sec5} have been obtained by assuming the unit-gain operation of all elements ($\mathbf{A}(t)=\mathbf{I}_N$); therefore, both of these modes do not influence the outcomes.
The per-element gain matrix is defined as
$\mathbf{A}(t) \triangleq \operatorname{diag}\!\left(\alpha_1(t),\dots,
\alpha_N(t)\right) \in \mathbb{R}^{N\times N}$, where $\alpha_n(t) \in \{0, 1, \alpha_{\mathrm{act}}\}$ reflects the operating mode of element $n$ at coherence interval $t$.
When all elements operate in unit-gain mode ($\alpha_n(t) = 1,\ \forall n$), 
$\mathbf{A}(t) = \mathbf{I}_N$, recovering the standard passive-RIS model.
The effective uplink channel vector from UE $k$ to AP $q$ is expressed in \eqref{eq:g_qk}.
\begin{figure*}[!t]
\vspace{1em}
\begin{equation}
\mathbf{g}_{q,k}(\boldsymbol{\phi}(t), \boldsymbol{\alpha}(t)) = \mathbf{H}_{{q}}(t)\,\mathbf{A}(t)\!
\left(\mathbf{\Psi}^\mathsf{r}(t)\, \mathbf{h}_{{k}}^{\mathsf{r}}(t) + 
\mathbf{\Psi}^\mathsf{t}(t)\, \mathbf{h}_{{k}}^{\mathsf{t}}(t)\right) \in \mathbb{C}^{Q\times 1}
\label{eq:g_qk}
\end{equation}
\vspace{-0.5em}
\hrule
\end{figure*}

\sloppy For notational brevity, $\mathbf{g}_{q,k}(\boldsymbol{\phi}(t), \boldsymbol{\alpha}(t)), \mathrm{SINR}_k(\boldsymbol{\phi}(t), \boldsymbol{\alpha}(t))$, 
and $\mathcal{R}_k(\boldsymbol{\phi}(t), \boldsymbol{\alpha}(t))$ are hereafter written with $\boldsymbol{\alpha}(t)$ implied throughout.
\vspace{-0.6em}
\subsection{Stateful Surface Model}
\vspace{-0.1em}
Conventional STAR-RIS models assume that the surface is memoryless. The phase configuration $\boldsymbol{\phi}(t)$ is set independently at each coherence interval in response to a control command issued by the AP, with no dependence on the surface state at  the prior time instance $t-1$. This ignores two physical realities. First, practical phase-shifting circuits (varactor diodes, PIN diodes, and liquid-crystal elements) exhibit settling transients, where the phase at time $t$ is a smoothed function of the current and previous phase states, not an instantaneous jump. Second, the control channel between the AP and the surface is bandwidth-limited, so transmitting a complete $2N$-dimensional update every coherence interval is costly.
We therefore introduce a stateful STAR-RIS model in which 
$\boldsymbol{\phi}(t)$, defined in~\eqref{eq:phi_def}, evolves 
according to temporal dynamics as
\vspace{-0.5em}
\begin{equation}
\boldsymbol{\phi}(t) = \beta\,\boldsymbol{\phi}(t-1) + (1-\beta)\,\boldsymbol{u}(t),
\label{eq:stateful_ris}
\end{equation}
where $\boldsymbol{u}(t) \in \mathbb{R}^{2N}$ is the control command issued by the meta-policy at coherence interval $t$, and $\beta \in [0,1)$ is the surface inertia coefficient that governs the inertia of the surface state. The value when $\beta = 0$ represents the conventional memoryless model, while $\beta \to 1$ corresponds to a surface that changes very slowly and relies almost entirely on its previous state.
Just as $\rho_c$ quantifies the temporal persistence of the channel, $\beta$ quantifies the temporal persistence of the surface configuration. When the channel evolves slowly ($\rho_c \approx 1$), the optimal phase vector changes little between intervals, so a surface with high $\beta$ can track the optimum with smaller control updates. Conversely, when the channel changes rapidly ($\rho_c \ll 1$), a lower $\beta$ is needed to allow fast adaptation. The optimal $\beta$ is therefore a function of $\rho_c$.
\vspace{-0.6em}
\subsection{Signal Model}
The received signal vector at $q^{th}$ AP  is defined as
\vspace{-0.7em}
\begin{equation}
\mathbf{y}_q(t) = \sum_{k=1}^{K} \mathbf{g}_{q,k}(\boldsymbol{\phi}(t)) 
\sqrt{p_k}\, x_k + \mathbf{n}_q(t),
\label{eq:received}
\end{equation}
where $p_k$ is the transmit power of UE $k$, $x_k$ is the unit-power transmitted symbol, and $\mathbf{n}_q(t) \sim \mathcal{CN}(\mathbf{0}, \sigma^2 \mathbf{I}_Q)$.
By employing maximum-ratio combining (MRC) at the $q^{th}$ AP \cite{ngo2013energy}, the SINR of UE $k$ at AP $q$ is
\begin{equation}
{\mathrm{SINR}_{k,q}}(\boldsymbol{\phi}(t)) =  
\frac{p_k \|\mathbf{g}_{q,k}(\boldsymbol{\phi}(t))\|_2^2}
{\displaystyle\sum_{i \neq k} p_i \!\left|
\frac{\mathbf{g}_{q,k}^H(\boldsymbol{\phi}(t))\, 
\mathbf{g}_{q,i}(\boldsymbol{\phi}(t))}
{\|\mathbf{g}_{q,k}(\boldsymbol{\phi}(t))\|_2}
\right|^2 + \sigma^2}.
\label{eq:SINR}
\end{equation}
Each UE $k$ is served by one AP $q_k$, and $\mathrm{SINR}_k(\boldsymbol{\phi}(t))\triangleq\mathrm{SINR}_{k,q_k}(\boldsymbol{\phi}(t))$. The $M$ APs share one STAR-RIS configuration. Combining the per-AP signals at a central processing unit, as in cell-free massive multiple-input multiple-output (MIMO) processing, is a direct extension that replaces $\mathrm{SINR}_{k,q_k}$ by an aggregate over $q$. All UEs are served by the same AP ($q_k=1$), consequently, the achievable uplink spectral efficiency per UE  
is expressed as $\mathcal{R}_k(\boldsymbol{\phi}(t)) = 
\log_2(1 + \mathrm{SINR}_k(\boldsymbol{\phi}(t)))$ 
[bps/Hz], and the sum-rate is 
$\mathcal{R}_{\mathrm{sum}}(t) = \sum_{k=1}^K 
\mathcal{R}_k(\boldsymbol{\phi}(t))$ [bps/Hz].
Our model notations are provided in Table~\ref{tab:parameter_definitions}.
\begin{table}[htbp]
\centering
\caption{Key Notations}
\label{tab:parameter_definitions}
\begin{tabular}{@{}p{1.8cm}p{6cm}@{}}
\toprule
\textbf{Symbol} & \textbf{Definition} \\
\midrule
$M, Q, K, N$ & APs, antennas/AP, UEs, RIS elements \\
$v_k, f_{D,k}, \rho_c$ & UE velocity, Doppler shift, temporal correlation \\
$\mathbf{H}_{{q}}(t)$ & RIS-to-AP channel matrix \\
{$\eta_n$} & {Energy-splitting ratio of element $n$} \\
$\mathbf{h}_{{k}}^{\mathsf{r}/\mathsf{t}}(t)$ & UE-to-RIS channel (reflection/transmission) \\
$\mathbf{g}_{q,k}(\boldsymbol{\phi}(t), \boldsymbol{\alpha}(t))$ & Effective uplink channel \\
$p_k, \sigma^2$ & UE transmit power, noise power \\
$\boldsymbol{\phi}^\mathsf{r}(t), \boldsymbol{\phi}^\mathsf{t}(t)$ & STAR-RIS reflection/transmission phases \\
$\mathbf{\Psi}^\mathsf{r}(t), \mathbf{\Psi}^\mathsf{t}(t)$ & STAR-RIS phase shift matrices \\
$\beta, \boldsymbol{u}(t)$ & Surface {inertia} coefficient, control command \\
$\beta_n, \lambda$ & Warm start element inertia, element memory decay  \\
$\boldsymbol{\alpha}(t), \mathbf{A}(t)$ & Per-element gain vector, diagonal gain matrix \\
$D_s, D_C$ & Scenario feature dimension, context vector dimension \\
$w_i(\tau), T_w$ & Episode relevance score, decay window \\

$T_E, \eta_{\mathcal{W}}$ & Capacity of the episodic memory, Step size of the tag-weight update \\
\bottomrule
\end{tabular}
\end{table}
\vspace{-0.9em}
\section{Problem Formulation and Optimization Problem}
\label{sec3}
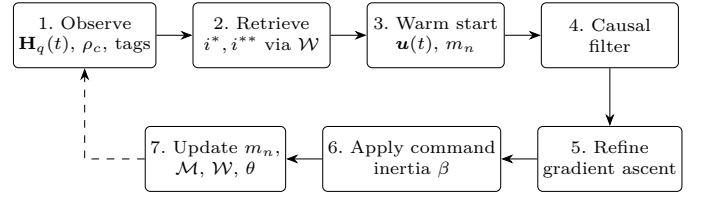
\begin{figure}[t]
\centering
\resizebox{\columnwidth}{!}{%
\begin{tikzpicture}[
    box/.style={
        draw=black,
        rounded corners=2pt,
        align=center,
        font=\scriptsize,
        minimum height=9mm,
        minimum width=19mm,
        inner sep=2pt
    },
    >=Stealth,
    node distance=5mm
]

\node[box] (obs) {1. Observe\\ $\mathbf{H}_q(t)$, $\rho_c$, tags};
\node[box,right=of obs] (ret) {2. Retrieve\\ $i^*,i^{**}$ via $\mathcal{W}$};
\node[box,right=of ret] (ws) {3. Warm start\\ $\boldsymbol{u}(t)$, $m_n$};
\node[box,right=of ws] (cf) {4. Causal\\ filter};

\node[box,below=8mm of cf] (gr) {5. Refine\\ gradient ascent};
\node[box,left=of gr] (st) {6. Apply command\\ inertia $\beta$};
\node[box,left=of st] (mem) {7. Update $m_n$,\\ $\mathcal{M}$, $\mathcal{W}$, $\theta$};

\draw[->,black] (obs)--(ret);
\draw[->,black] (ret)--(ws);
\draw[->,black] (ws)--(cf);
\draw[->,black] (cf)--(gr);
\draw[->,black] (gr)--(st);
\draw[->,black] (st)--(mem);
\draw[->,dashed,black] (mem.west) -| (obs.south);

\end{tikzpicture}}
\caption{Meta2L pipeline, executed once per coherence interval. Steps 1 to 6 form the fast loop within an interval; the dashed arrow returns to step 1 at the next interval.}
\label{fig:pipeline}
\end{figure}

In current RIS designs, the surface is completely memoryless and does not keep track of any past settings, and cannot track how the channel has been evolving over time.
Each coherence interval triggers a full re-optimization from a cold start, increasing computational costs and signaling overhead.
Memoryless models lack persistent state across coherence intervals and thus cannot accumulate experience as defined in the present paper.
Meta2L runs the loop of Fig.~\ref{fig:pipeline} once per coherence interval.  (1) The controller monitors the channel and computes $\rho_c$ as well as the tag vector for the current context. (2) Then, it selects the two most relevant past episodes in $\mathcal{M}$ using the tag weights $\mathcal{W}$ (Section~\ref{sec:omd}). (3) The system then initializes the phases using the weighted average of the Doppler shifts in the selected episodes and in the element memory (Section~\ref{sec3}-C). (4) The causal filter filters out physically inconsistent initializations (Section~\ref{sec:cif}). (5) The projected gradient ascent method improves the phases. (6) The controller sends the command to the surface where it will be processed according to its inertial dynamics with parameter $\beta$, and $\beta$ is updated. (7) The element memory, episodic memory, tag weights and meta-policy are updated. Steps (1) to (6) form the fast loop within a coherence interval while step (7) is the slow loop over intervals. The rest of the section provides the details of each of the blocks. 
The update (\ref{eq:element_memory}) and the initialization (\ref{eq:memory_phase}) of element $n$ use only the history of element $n$. Keeping the same $N$ states in a controller table will result in having the same initial state for that table, and using a centralized approach will yield the same sum rate and gradient steps. This was verified numerically and found that the centralized variant Meta2L-C reproduces every result of Meta2L (Table~\ref{tab:reuse}). The distribution brings no performance gain. What it changes is the architecture, where the per-element state does not have to be stored or read back by the controller. The key contribution is the per-element state carried across coherence intervals, rather than where that state is stored.
The surface inertia $\beta$ is the physical part of the model as it represents the settling of the phase-shifting circuit, and a controller can also impose it digitally. The element memory $m_n$ is a logical per-element state that may be held in the driver of each unit cell or, equivalently for the algorithm, in a controller table with one entry per element. The episodic memory $\mathcal{M}$ is always held at the controller.
This work proposes integrating two memory types to turn the surface into a stateful surface with element-level memory: (i) a surface inertia coefficient $\beta$, which models the settling transients and temporal inertia. It assumes that the realized phase is a smoothed version of the previous state and the control command.
(ii) a per-element episodic memory that integrates each element's phase history across episodes using an exponential moving average. This memory is updated at the same rate as the model. This per-element memory is unique to each element and reflects the spatial structure of the channel. 

Phase-control power consumption varies based on the operating modes. For passive devices, power consumption comes from the PIN/varactor diode and is proportional to the square of the phase difference. For semi-active components, additional amplification requires power from an external energy source; in the case of sleep mode elements, there is no associated energy cost.
\vspace{-0.8em}
\subsection{Surface Inertia (Stateful Surface Model)}
For each coherence interval, the controller determines which control command should be issued to the surface. Under the stateful model, the optimization variables are the control command $\boldsymbol{u}(t) \in \mathbb{R}^{2N}$ and the per-element mode selection vector $\boldsymbol{\alpha}(t) \in \{0,1,\alpha_{\mathrm{act}}\}^N$. The actual phase $\boldsymbol{\phi}(t)$ is not the optimization variable; it appears naturally through the surface dynamics.
The objective balances spectral efficiency, phase control cost, control smoothness, and RIS power consumption.
We formulate the real-time intra-episode stateful optimization problem as follows:
\vspace{-1em}
\begin{subequations}
\label{eq:weighted_stateful}
\begin{align}
\max_{\boldsymbol{u}(t),\,\boldsymbol{\alpha}(t)} \; 
F_\beta(\boldsymbol{u},\boldsymbol{\alpha}) 
&= \omega_1 \sum_{k=1}^K 
   \mathcal{R}_k(\boldsymbol{\phi}(t)) \nonumber\\[-0.5em]
&\quad - \omega_2 \sum_{n=1}^N c_n
         \left[(\phi_n^\mathsf{r}(t))^2+(\phi_n^{\mathsf{t}}(t))^2\right]
   \nonumber\\[-0.5em]
&\quad - \lambda_c\|\boldsymbol{u} - \boldsymbol{\phi}(t-1)\|_2^2 \nonumber\\[-0.5em]
&\quad - \lambda_{\mathrm{ris}}\sum_{n=1}^N P_n(\alpha_n(t)),
\label{eq:weighted_stateful_a} \\[-0.2em]
\text{s.t.} \quad 
& u_n^{\mathsf{r}},\, u_n^{\mathsf{t}} \in [0, 2\pi), \quad \forall\, n,
\label{eq:weighted_stateful_b} \\[-0.2em]
& \alpha_n(t) \in \{0,\, 1,\, \alpha_{\mathrm{act}}\}, \quad \forall\, n,
\label{eq:weighted_stateful_c}
\end{align}
\end{subequations}
where $\omega_1$ is the spectral efficiency coefficient, $\omega_2$ is the phase control energy coefficient, and $c_n = \mathbf{1}[\alpha_n(t)]$ is the element activity indicator. 
The quadratic term $(\phi_n^{\mathsf{r}}(t))^2 + (\phi_n^{\mathsf{t}}(t))^2$ results in an approximation to the reactive control energy consumed around the operating point. $\lambda_c$ is the control innovation penalty, which punishes for large deviations of $\boldsymbol{u}(t)$ from the previous 
    surface state $\boldsymbol{\phi}(t-1)$,
$\lambda_{\mathrm{ris}}$ is the RIS hardware energy penalty, $P_n(\alpha_{n}(t)) \in \{0,\, P_{\mathrm{pass}},\, 
    P_{\mathrm{act}}\}$ is the per-element power consumption in sleep, passive, and semi-active mode, respectively, and 
$P_{\mathrm{RIS}}(t) = \sum_{n=1}^N P_n(\alpha_n(t))$ is the total RIS power. 
The constraint~\eqref{eq:weighted_stateful_b} restricts each element's control command to be within the physically acceptable phase range. The constraint~\eqref{eq:weighted_stateful_c} allows each element to operate in one operating mode.
The discreteness of $\boldsymbol{\alpha}$ is addressed by fixing the mode selection at the 
meta-policy level by using a greedy energy-rate 
trade-off. Element $n$ is set to sleep mode 
($\alpha_n(t){=}0$) if $\max_k|\partial\,\mathrm{SINR}_k(\boldsymbol{\phi}(t))/\partial\phi_n^\mathsf{r}|$ 
goes below $\theta_l$, to semi-active 
mode ($\alpha_n(t){=}\alpha_{\mathrm{act}}$) in case it 
goes above $\theta_h$, and to passive mode otherwise. 
Then, the variable $\boldsymbol{u}$ is optimized 
assuming $\boldsymbol{\alpha}$ is fixed.
For the specific scenario where there are no surface dynamics ($\beta{=}0$), the phase becomes the control variable itself.
The transition from the parameter $\boldsymbol{u}$ to $\boldsymbol{\phi}$ represents this simplification since in the memoryless case, the control input and the actual phase become the same.
The memoryless objective, which optimizes directly over the realized phase $\boldsymbol{\phi}$
and mode vector $\boldsymbol{\alpha}$
with no surface dynamics is expressed as
\vspace{-0.7em}
\begin{subequations}
\label{eq:weighted}
\begin{align}
\max_{\boldsymbol{\phi},\,\boldsymbol{\alpha}} \; 
F(\boldsymbol{\phi},\boldsymbol{\alpha}) 
&= \omega_1 \sum_{k=1}^K \mathcal{R}_k(\boldsymbol{\phi}(t))
\nonumber\\[-0.3em]
&\quad - \omega_2 \sum_{n=1}^N c_n\left[(\phi_n^{\mathsf{r}})^2+(\phi_n^{\mathsf{t}})^2\right]\nonumber
\\[-0.3em]
&\quad - \lambda_{\mathrm{ris}}
  \sum_{n=1}^N P_n(\alpha_n(t)), \nonumber\\[-0.5em]
\label{eq:weighted_a} \\[-0.1em]
\text{s.t.} \quad 
& \phi_n^{\mathsf{r}}, \phi_n^{\mathsf{t}} \in [0, 2\pi), \quad \forall\, n,
\label{eq:weighted_b} \\[-0.3em]
& \alpha_n(t) \in \{0,\, 1,\, \alpha_{\mathrm{act}}\}, \quad \forall\, n.
\label{eq:weighted_c}
\end{align}
\end{subequations}
When we substitute with the values of $\beta = 0$, $\boldsymbol{u} = \boldsymbol{\phi}$, and 
$\lambda_c = 0$ into \eqref{eq:weighted_stateful}, the stateful objective collapses back to the memoryless form in \eqref{eq:weighted}.
A high value of the surface inertia coefficient $\beta$ smooths the phase and reduces control overhead; however, it causes a lag when the channel changes rapidly, and the opposite takes place with a low value. The optimal value $\beta^*$ that will minimize the loss of rate due to the tracking error is derived in Appendix D. Every command issued per interval is generated through a finite number of iterations using gradients, which leaves it with an error variance of $\sigma_w^2$. The surface inertia smooths out the residual but is slow to respond compared to the channel, and the net effect yields the closed form (\ref{eq:beta_opt}). In that case, $\beta^*$ will depend solely on $\rho_c$, as well as on the ratio $C_\Delta^2/\sigma_w^2$; the smoothness parameter $\beta_F$ will scale both. As such, $\beta^*$ increases with $\rho_c$ as a slowly varying channel lets the surface average over more intervals, and the control penalty $\lambda_c$ in (\ref{eq:weighted_stateful}) raises it further.
The closed-form expression is used only as a reference, and the controller does not evaluate it. Instead, it learns $\beta$ online by minimizing the per-interval cost as 
$J_t(\beta)=\omega_1[\mathcal{R}_{\mathrm{sum}}(\boldsymbol{u}(t))-\mathcal{R}_{\mathrm{sum}}(\boldsymbol{\phi}(t))]+\lambda_c\|\boldsymbol{\phi}(t)-\boldsymbol{\phi}(t-1)\|_2^2$,
which captures the rate loss due to inertia together with the control innovation in (\ref{eq:weighted_stateful}). Differentiating (\ref{eq:stateful_ris}) with respect to time gives the exact recursion as follows:
\begin{subequations}\label{eq:beta_update}
\begin{align}
&{\mathbf{s}(t)=\boldsymbol{\phi}(t-1)-\boldsymbol{u}(t)+\beta\,\mathbf{s}(t-1),}\\
&{\frac{\partial J_t}{\partial\beta}=-\omega_1\nabla\mathcal{R}_{\mathrm{sum}}(\boldsymbol{\phi}(t))^{T}\mathbf{s}(t)}\nonumber\\
&\qquad+2\lambda_c\big(\boldsymbol{\phi}(t)-\boldsymbol{\phi}(t-1)\big)^{T}\big(\mathbf{s}(t)-\mathbf{s}(t-1)\big),
\end{align}
\end{subequations}
{where $\mathbf{s}(t)=\partial\boldsymbol{\phi}(t)/\partial\beta$, phase differences are wrapped to $[-\pi,\pi)$, and $\beta$ is updated by an adaptive gradient step on its logit. The update uses only the measured rate, the SINR gradient already computed by the inner loop, and the operator's own weight $\lambda_c$. It needs neither $\beta_F$ nor the channel statistics $\rho_c$, $C_\Delta$ and $\sigma_w$. This is the precise sense in which the surface tunes its inertia without knowing the constants in (\ref{eq:beta_opt}); Fig.~\ref{fig5} compares the learned values with (\ref{eq:beta_opt}) and with an exhaustive-search optimum.}
The surface memory influences the effective channel statistics, so the expected stateful signal power becomes \eqref{eq:expected_power_stateful}.
\begin{figure*}[!t]
\vspace{0.5em}
\begin{align}
\mathbb{E}\!\left[\|\mathbf{g}_{q,k}(t)\|_2^2\right] &= \mathbb{E}\!\left[\|\mathbf{g}_{q,k}^{\mathsf{r}}(t) + \mathbf{g}_{q,k}^{\mathsf{t}}(t)\|_2^2\right] 
= S_k^{\mathsf{r}}(\beta,\rho_c) + S_k^{\mathsf{t}}(\beta,\rho_c) + 2\Re\!\left\{S_k^{(\times)}(\beta,\rho_c)\right\}
\label{eq:expected_power_stateful}
\end{align}
\vspace{-0.5em}
\hrule
\end{figure*}
where the reflection, transmission, and cross-terms now depend on both $\beta$ and $\rho_c$ through the stateful phase covariance as follows:
\begin{equation}
\boldsymbol{\Sigma}_\phi \triangleq \mathbb{E}\!\left[\boldsymbol{\phi}(t)\boldsymbol{\phi}^T(t)\right] = \frac{(1-\beta)^2}{1-\beta^2}\,\mathbb{E}\!\left[\boldsymbol{u}(t)\boldsymbol{u}^T(t)\right] + \frac{\beta^2}{1-\beta^2}\boldsymbol{\Sigma}_e,
\label{eq:phase_covariance}
\end{equation}
where $\boldsymbol{\Sigma}_e = \mathbb{E}[\boldsymbol{e}(t)\boldsymbol{e}^T(t)]$ is the tracking error covariance. The full derivation that accounts for the temporal correlation introduced by $\beta$ can be found in Appendix~A. 
\vspace{-0.8em}
\subsection{Element Local Memory}
Each RIS element keeps its own local phase history as a logical per-element state, so historical information is distributed across the $N$ elements instead of being kept in one controller database. The element-memory vector $\mathbf{m}(t) = [m_1(t),\dots,m_N(t)]^T$
offloads the role of storing local phase history away from a centralized database. It collects the $N$ element states and is distinct from the surface inertia coefficient $\beta$, which is a single scalar shared by all elements.

Each element $n$ maintains a local memory state $m_n(t) \in \mathbb{R}$ that accumulates its own phase history via an exponential moving average, updated once per episode after gradient convergence as follows:
\vspace{-0.4em}
\begin{equation}
m_n(t) = \lambda\, m_n(t-1) + (1-\lambda)\,\phi_n^*(t),
\label{eq:element_memory}
\end{equation}
\vspace{0.2em}
where $\phi_n^*(t)$ is the converged optimal phase of element $n$ at episode $t$, and $\lambda \in [0,1)$ is the \emph{local memory decay coefficient} governing how quickly the historical influence fades. A high $\lambda$ retains longer phase history, while $\lambda \to 0$ reduces $m_n(t)$ to the current phase. 
An episodic memory $\mathcal{M}$ stores scenario features, compressed tags, and optimal phase vectors for warm-start retrieval, while element-level phase history is handled locally on the surface.
The episode memory $\mathcal{M}$ contains $T_{E}$ raw episodes and includes the following components:
\vspace{-0.4em}
\begin{equation}
\mathcal{M} = \left\{\bigl(\mathbf{s}_i,\, \tilde{\mathbf{t}}_i,\, 
\mathbf{a}_i,\, w_i(\tau)\bigr)\right\}_{i=1}^{T_{E}},
\end{equation}
where $\mathbf{s}_i \in \mathbb{R}^{D_s}$ ($D_s{=}64$) is the scenario features vector 
comprising instantaneous channel statistics, user positions, velocity estimates, 
Doppler spread, AP load, and interference level. Also, $\tilde{\mathbf{t}}_i \in \{0,1\}^{D_t}$ is the compressed tag 
vector that is produced by the autoencoder and generates interpretable episode 
summaries that are used in the similarity search. 
$\mathbf{a}_i = (\boldsymbol{\phi}^{\mathsf{r}*}, 
\boldsymbol{\phi}^{\mathsf{t}*}) \in [0,2\pi)^{2N}$ encodes the optimal phase shift 
vectors retrieved during warm-start; and $w_i(\tau) = \exp(-\tau/T_w)$ 
is the time-dependent relevance score that weights the contribution of each 
past episode to the similarity search, where $\tau = t - i$ is the episode age and $T_w$ is a decay window hyperparameter. Per-element local phase 
history is not stored in $\mathcal{M}$ but is instead maintained 
distributedly within the surface itself via the element memory states $\{m_n(t)\}$ defined in~\eqref{eq:element_memory}, which are updated once per episode.

Each STAR-RIS scenario is characterized by 
a tag vector for episode $i$ as $\mathbf{t}_i \in \{0,1\}^{D_t}$ 
that encodes environmental, RIS, network, 
and QoS attributes, such as path loss, Doppler effect, aperture size, phase control, and energy budget. 
Each tag vector $\mathbf{t}_i$ 
is stored directly as a $D_t$-dimensional 
binary vector within the episodic tuple 
in memory $\mathcal{M}$. The compressed form  
representation $\tilde{\mathbf{t}}_i$ that is
adopted by the meta-policy is generated 
using a shallow autoencoder. 
The binary bottleneck is generated via a straight-through estimator that is trained offline on the 
collected episode tags to minimize binary cross-entropy reconstruction loss. This 
weighting scheme fulfills the goal of transparency in the tag design and regret analysis 
of Theorem~\ref{thm:regret}.
\vspace{-0.5em}
\subsection{Doppler-Aware Warm-Start Initialization}
Let $i^*$ and $i^{**}$ be the episode indices, having the highest similarity scores for the current scenario in the first memory stage $\mathcal{M}$.
We define the warm-start phase vector as
\begin{equation}
\boldsymbol{\phi}^{(0)}(t) = \rho_c \boldsymbol{\phi}^*_{i^*} + (1-\rho_c)\boldsymbol{\phi}^*_{i^{**}},
\label{eq:warmstart}
\end{equation}
where $\rho_c = J_0(2\pi f_{D,\max} T_s)$ ensures that the warm-start gap $\Delta_{\rho_c}$ is conservative across all UEs; a high $\rho_c$ places more weight on the closest-match episode since the channel has changed little.
The two memory mechanisms are architecturally distinct. The surface inertia coefficient $\beta$ models hardware settling dynamics that affect all elements identically and acts once per coherence interval on the converged command, as in (\ref{eq:stateful_ris}). The per-element memory state $m_n(t)$, by contrast, operates at the episode timescale and is updated once per coherence interval, after the intra-episode gradient optimization has converged, and it provides element-specific historical context that varies across the array. This timescale separation ensures that $m_n(t)$ accumulates cross-episode history rather than intra-episode transients, making the two mechanisms strictly complementary rather than redundant.

The local memory state $m_n(t)$ enters the warm-start initialization for the next episode as
\vspace{-0.2em}
\begin{equation}
\phi_n^{(0)}(t) = \beta_n\,\phi_n^*(t-1) + 
(1-\beta_n)\,u_n(t) + \gamma_n\,m_n(t-1),
\label{eq:memory_phase}
\end{equation}
where $u_n(t)$ is the $n$-th component of the warm-start vector $\boldsymbol{\phi}^{(0)}(t)$, $\beta_n \in [0,1)$ 
is the per-element inertia coefficient, and $\gamma_n$ is the local memory gain that scales the influence of the accumulated 
history on the initialization point. The same equation applies separately to $\phi_n^{\mathsf{r}}(t)$ and $\phi_n^{\mathsf{t}}(t)$. When $\gamma_n = 0,\ \forall\, n$, equation~\eqref{eq:memory_phase} reduces to the element-wise form of~\eqref{eq:stateful_ris} evaluated at the warm-start point. 
This does not merely follow a command from the controller but modulates its response based on what phases it has previously realized. Collectively, the $N$ elements hold a logically distributed memory in which the initialization of each element depends only on its own history and on the retrieved command, without any inter-element communication. This is conceptually analogous to a memristive or neuromorphic surface, where local state variables participate in computation, and it moves the surface from a passive infrastructure element toward a stateful, adaptive component of the radio environment.
\vspace{-0.8em}
\subsection{Phase Stationarity Conditions}
\begin{proof2}
We derive the stationarity conditions for the element phase optimization. The Lagrangian for the phase constraint $\phi_n \in [0, 2\pi)$ is expressed as
\begin{align}
L &= \omega_1 \sum_{k=1}^K 
\mathcal{R}_k(\boldsymbol{\phi}(t),
\boldsymbol{\alpha}(t))
- \omega_2 \sum_{n=1}^N c_n
\left[(\phi_n^{\mathsf{r}}(t))^2
+(\phi_n^{\mathsf{t}}(t))^2\right] 
\nonumber\\
&\quad - \sum_{n=1}^N 
\mu_n^{\mathsf{r},+}
(\phi_n^{\mathsf{r}}(t) - 2\pi) 
- \sum_{n=1}^N 
\mu_n^{\mathsf{t},+}
(\phi_n^{\mathsf{t}}(t) - 2\pi)
\nonumber\\
&\quad + \sum_{n=1}^N 
\mu_n^{\mathsf{r},-}\phi_n^{\mathsf{r}}(t)
+ \sum_{n=1}^N 
\mu_n^{\mathsf{t},-}\phi_n^{\mathsf{t}}(t).
\label{lagrangian}
\end{align}
Using the chain rule, we obtain 
\begin{equation}
\frac{\partial \mathcal{R}_k(\boldsymbol{\phi}(t))}{\partial \phi_n^\mathsf{r}} = \frac{1}{\ln 2}\frac{1}{1+\mathrm{SINR}_k(\boldsymbol{\phi}(t))}\frac{\partial\,\mathrm{SINR}_k(\boldsymbol{\phi}(t))}{\partial\phi_n^\mathsf{r}},
\end{equation}
where $G_k \triangleq G_k(\boldsymbol{\phi}(t)) 
= \|\mathbf{g}_{q,k}(\boldsymbol{\phi}(t),
\boldsymbol{\alpha}(t))\|_2^2$ and 
$I_k \triangleq \sum_{i\neq k} p_i \left|\frac{\mathbf{g}_{q,k}^H \mathbf{g}_{q,i}}{\|\mathbf{g}_{q,k}\|_2}\right|^2 + \sigma^2$.
Let $\mathbf{c}_{q,k} \triangleq
\mathbf{H}_{{q}}^H(t)\,
\mathbf{g}_{q,k} \in \mathbb{C}^{N\times 1}$, {with $q=q_k$,} then the signal power gradient is expressed as
\begin{equation}
{\frac{\partial G_k}{\partial \phi_n^\mathsf{r}}
= 2\Re\!\left\{j\,\alpha_n\sqrt{\eta_n}\,e^{j\phi_n^{\mathsf{r}}(t)}\,
[\mathbf{h}_{k}^\mathsf{r}(t)]_n\,
[\mathbf{c}_{q,k}]_n^{*}\right\}.}
\label{eq:dGk}
\end{equation}
Hence, the SINR gradient with respect to $\phi_n^\mathsf{r}$ is expressed as
\begin{equation}
\frac{\partial\,\mathrm{SINR}_k}{\partial\phi_n^\mathsf{r}} =\!
\frac{p_k}{I_k}\!
\frac{\partial G_k}{\partial\phi_n^\mathsf{r}}\! 
-\! \frac{p_k G_k}{I_k^2}\!
\frac{\partial I_k}{\partial\phi_n^\mathsf{r}}.
\label{eq:dSINR_dphi}
\end{equation}
Setting $\partial L/\partial \phi_n^\mathsf{r} = 0$ yields
\begin{align}
\frac{\omega_1}{\ln 2}\sum_{k=1}^K 
\frac{1}{1+\mathrm{SINR}_k}
\frac{\partial\,\mathrm{SINR}_k}{\partial\phi_n^\mathsf{r}}\! 
-\! 2\omega_2 c_n \phi_n^\mathsf{r}\! 
+\! \mu_n^{\mathsf{r},-}\! -\! \mu_n^{\mathsf{r},+}\! =\! 0,
\label{eq:stationarity}
\end{align}
where $\mu_n^{\mathsf{r},+} \geq 0$ and $\mu_n^{\mathsf{r},-} \geq 0$ 
are the KKT multipliers associated with the 
upper bound $\phi_n^\mathsf{r} \leq 2\pi$ and lower 
bound $\phi_n^\mathsf{r} \geq 0$ constraints 
respectively, and all terms are evaluated 
at $\boldsymbol{\phi}(t)$. An analogous condition holds for $\phi_n^\mathsf{t}$ 
by replacing the superscript $\mathsf{r}$ with $\mathsf{t}$ throughout.
\end{proof2}
\begin{proposition}
If the constraint is inactive ($\mu_n^{\mathsf{r},+} = \mu_n^{\mathsf{r},-} = 0$) and element $n$ is active ($c_n=1$), the stationary-point condition reduces to
\vspace{-0.em}
\begin{equation}
\phi_n^{\mathsf{r}*} = \frac{\omega_1}{2\omega_2\ln 2}\sum_{k=1}^K \frac{1}{1+\mathrm{SINR}_k(\boldsymbol{\phi}^*(t))}\frac{\partial\,\mathrm{SINR}_k(\boldsymbol{\phi}^*(t))}{\partial\phi_n^\mathsf{r}}.
\label{eq:interior_sol}
\end{equation}
For multiple UEs, this is solved via successive substitution.
If either bound is active, the closed-form projection solution is expressed as
\begin{equation}
\phi_n^{\mathsf{r}*} = \operatorname{clip}\!\left(
\frac{\omega_1}{2\omega_2\ln 2}
\sum_{k=1}^K 
\frac{1}{1+\mathrm{SINR}_k}
\frac{\partial\,\mathrm{SINR}_k}
{\partial\phi_n^\mathsf{r}},\; 0,\; 2\pi\right),
\label{eq:boundary_sol}
\end{equation}
where all terms are evaluated at 
$\boldsymbol{\phi}^*(t)$, and 
$\operatorname{clip}(x, 0, 2\pi)$ projects 
$x$ onto $[0, 2\pi)$,
applied separately to $\phi_n^{\mathsf{r}}(t)$ and $\phi_n^{\mathsf{t}}(t)$ 
by replacing $\mathsf{r}$ with $\mathsf{t}$ throughout.
\end{proposition}
All theoretical results of this section are stated for a fixed mode vector $\boldsymbol{\alpha}$ and rely on the assumptions below. Each result states the assumptions it uses. None is claimed for arbitrary channel processes or arbitrary hardware. {(A1) The objective $F(\cdot,\boldsymbol{\alpha})$ is $\beta_F$-smooth on the feasible set. (A2) The initialization ensures that $\|\boldsymbol{\phi}^{(0)}(t)-\boldsymbol{\phi}^*(t)\|_2\le\Delta_{\rho_c}=C_\Delta\sqrt{1-\rho_c^2}$ for a finite constant $C_\Delta$; the latter inequality is inspired from (\ref{eq:gauss_markov}), yet is stated instead of being proved. (A3) The vector $\boldsymbol{\phi}^*(t)$ is a stationary point of the nonconvex optimization problem $F$, which is in general a local maximum, not a global one, and such that $F(\boldsymbol{\phi}^{(s)})\le F(\boldsymbol{\phi}^*(t))$ and $F(\boldsymbol{\phi}^*)-F(\boldsymbol{\phi})\le\frac{\beta_F}{2}\|\boldsymbol{\phi}-\boldsymbol{\phi}^*\|_2^2$. (A4) The increments $\boldsymbol{\phi}^*(t)-\boldsymbol{\phi}^*(t-1)$ are uncorrelated over $t$ and satisfy $\mathbb{E}\|\boldsymbol{\phi}^*(t)-\boldsymbol{\phi}^*(t-1)\|_2^2\le C_\Delta^2(1-\rho_c^2)$.} {(A5) Once the phases are centered by their per-element long-run mean, the memory $m_n(t-1)$ is independent of the new channel innovation and $\mathrm{Cov}[\phi_n^*(t),\bar{m}_n]>0$, which is the case when the array has persistent per-element structure such as a static LoS component. (A6) All channels have the first order Gauss–Markov model (\ref{eq:gauss_markov}), where $\rho_c$ and the large scale parameters are constant throughout the analysis period. (A7) The phase shifter uses the stateful framework (\ref{eq:stateful_ris}), where the experiments in Section~\ref{sec5} use the following two mechanisms for settling and quantization: (1) the convex combination of the current and desired phase and (2) rounding to $2^b$ levels. Mutual coupling is not taken into account. (A8) The loss function due to the tag loss is linear in $\mathcal{W}$, $\|\nabla \ell_t \|_2 \leq L$, and independent of the current weight configuration.}
{Stationarity is measured by the gradient mapping $\mathcal{G}(\boldsymbol{\phi})\triangleq\beta_F\big(\Pi(\boldsymbol{\phi}+\nabla F(\boldsymbol{\phi})/\beta_F)-\boldsymbol{\phi}\big)$, where $\Pi$ is the projection onto $[0,2\pi)^{2N}$, and a point is $\epsilon_g$-stationary if $\|\mathcal{G}\|_2\le\epsilon_g$.}
\begin{theorem}[Doppler-Aware Warm-Start Convergence]
\label{thm:warmstart}
For A1-A3, the gradient ascent by using $1/\beta_F$ as a step size starting from $\boldsymbol{\phi}^{(0)}(t)$ is represented as
$\min_{s<T}\|\mathcal{G}(\boldsymbol{\phi}^{(s)})\|_2^2\le\beta_F^2\Delta_{\rho_c}^2/T$.
Thus, the number of steps required to reach an $\epsilon_g$-stationary point is obtained as
\vspace{-0.3em}
\begin{equation}
T_{\mathrm{grad}} = \frac{\beta_F^2(1-\rho_c^2)C_\Delta^2}{\epsilon_g^2},
\end{equation}
which is rounded up to the nearest integer. With random initialization, the corresponding bound is expressed as $\beta_F^2\|\boldsymbol{\phi}_{\max}\|_2^2/\epsilon_g^2$, where $\|\boldsymbol{\phi}_{\max}\|_2$ bounds the initial distance to $\boldsymbol{\phi}^*$. $\mathcal{O}(1/\epsilon_g^2)$ dependency is a common feature of smooth nonconvex functions \cite{nesterov2004introductory}; however, the warm start has an effect on the constant, and not on the asymptotic order.
\end{theorem}
\begin{proof2}[Proof of Theorem~\ref{thm:warmstart}]
When $\boldsymbol{\phi}$ is a $\beta_F$-smooth function with step size $1/\beta_F$, it is ensured that $F(\boldsymbol{\phi}^{(s+1)})\ge F(\boldsymbol{\phi}^{(s)})+\frac{1}{2\beta_F}\|\mathcal{G}(\boldsymbol{\phi}^{(s)})\|_2^2$ \cite{nesterov2004introductory}. When we sum over $s<T$ and using A3, we get
\vspace{-0.4em}
\begin{align}
{\sum_{s<T}\|\mathcal{G}(\boldsymbol{\phi}^{(s)})\|_2^2}
&\le 2\beta_F\big(F(\boldsymbol{\phi}^*)-F(\boldsymbol{\phi}^{(0)})\big)\nonumber\\[-0.5em]
&\le\beta_F^2\|\boldsymbol{\phi}^{(0)}-\boldsymbol{\phi}^*\|_2^2.
\label{eq:descent}
\end{align}
Plugging in A2, dividing by $T$, and insisting that the right hand side is bounded above by $\epsilon_g^2$ yields the aforementioned $T_{\mathrm{grad}}$. This theorem pertains only to the stationary point near which the warm-start occurs, and does not imply any form of global optimality.
\end{proof2}
If the channel is fully uncorrelated, the warm-start reduces to a random initialization and provides no benefit.
The AP transmits only the innovation $\boldsymbol{\delta}(t) \triangleq \boldsymbol{u}(t) - \boldsymbol{\phi}(t-1)$, whose squared norm scales as $\mathcal{O}(1-\rho_c^2)$, matching the warm-start gap $\Delta_{\rho_c}^2/\|\boldsymbol{\phi}_{\max}\|_2^2$ from Theorem~\ref{thm:warmstart}. 
Both the stateful surface and the warm-start initialization exploit the same Gauss--Markov structure from complementary directions: one at the hardware level and one at the algorithmic level.
\begin{theorem}
\label{thm:stateful_warmstart}
In the stateful STAR-RIS model~\eqref{eq:stateful_ris} with inertia parameter $\beta \in [0,1)$, suppose A1 to A4 hold, and the controller starts from the realization of the surface state. Therefore, the effective initialization gap satisfies the following:
\vspace{-0.5em}
\begin{equation}
\mathbb{E}\|\boldsymbol{\phi}^{(0)}(t) - \boldsymbol{\phi}^*(t)\|_2^2 \leq \Delta_{\rho_c,\beta}^2 \triangleq \frac{C_\Delta^2(1-\rho_c^2)}{1-\beta^2}.
\label{eq:stateful_gap}
\end{equation}
The projected gradient ascent from $\boldsymbol{\phi}^{(0)}(t)$ reaches an $\epsilon_g$-stationary point within the expected gradient steps, which are expressed as
\begin{equation}
T_{\mathrm{grad}}^\beta = \frac{\beta_F^2 C_\Delta^2(1-\rho_c^2)}{\epsilon_g^2(1-\beta^2)},
\end{equation}
where $\beta_F$ is the smoothness constant of $F(\boldsymbol{\phi}, \boldsymbol{\alpha})$. As $1/(1-\beta^2)\geq1$, the effect of inertia increases the estimate by a factor of $1/(1-\beta^2)$ compared to the perfect surface that achieves the target phase angle.
\end{theorem}
\begin{proof2}[Proof of Theorem~\ref{thm:stateful_warmstart}]
The covariance of the tracking error~\eqref{eq:tracking_error} adds another contribution of $\beta^2 C_\Delta^2(1-\rho_c^2)/(1-\beta^2)$ to the initialization gap squared, resulting in~\eqref{eq:stateful_gap}. A4 adds these terms together, and $C_\Delta^2(1-\rho_c^2)[1+\beta^2/(1-\beta^2)]=C_\Delta^2(1-\rho_c^2)/(1-\beta^2)$. Substituting $\Delta_{\rho_c}^2$ by $\Delta_{\rho_c,\beta}^2$ in (\ref{eq:descent}) leads to $T_{\mathrm{grad}}^\beta$.
\end{proof2}
{Theorem~\ref{thm:stateful_warmstart} states a cost of inertia on the step bound, not a gain. The benefit of $\beta$ lies elsewhere: it averages the residual error of the finite-iteration solver and it smooths the realized trajectory, since $\boldsymbol{\phi}(t)-\boldsymbol{\phi}(t-1)=(1-\beta)(\boldsymbol{u}(t)-\boldsymbol{\phi}(t-1))$. The learned $\beta$ trades the factor $1/(1-\beta^2)$ against these effects, which is why it stays small in fast channels (Fig.~\ref{fig5}).}
\begin{theorem}
\label{thm:element_memory}
Assuming A4 and A5, let $\boldsymbol{e}_\beta(t) \triangleq \boldsymbol{\phi}^*(t) - \boldsymbol{\phi}_{{\beta}}^{(0)}(t)$ denote the per-element initialization error under the stateful model alone  with $\gamma_n = 0$, and let $\boldsymbol{e}_m(t) \triangleq \boldsymbol{\phi}^*(t) - \boldsymbol{\phi}_{{m}}^{(0)}(t)$ denote the initialization error under the memory-augmented warm-start~\eqref{eq:memory_phase} with $\gamma_n > 0$. 
The steady-state per-element variance of the initialization error satisfies
\begin{equation}
\mathrm{Var}\!\left[e_{m,n}(t)\right] \leq \mathrm{Var}\!\left[e_{\beta,n}(t)\right] - {2\gamma_n(1-\lambda)\,\mathrm{Cov}\!\left[\phi_n^*(t),\, \bar{m}_n\right]},
\label{eq:variance_reduction}
\end{equation}
where $\bar{m}_n = \mathbb{E}[m_n(t)]$ is the steady-state mean of the local memory state. If $\mathrm{Cov}[\phi_n^*(t), \bar{m}_n] > 0$ (i.e., the element's historical mean phase is positively correlated with the current optimal phase, which holds whenever the channel exhibits persistent spatial structure), then
\begin{equation}
\mathrm{Var}\!\left[e_{m,n}(t)\right] < \mathrm{Var}\!\left[e_{\beta,n}(t)\right], \quad {0<\gamma_n<\bar{\gamma}_n,}
\label{eq:strict_reduction}
\end{equation}
\vspace{0.2em}
where $\bar{\gamma}_n\triangleq 2|\mathrm{Cov}[e_{\beta,n},m_n]|/\mathrm{Var}[m_n]$.
Consequently, the memory-augmented initialization gap satisfies $\mathbb{E}[\|e_{m,n}\|^2] < \mathbb{E}[\|e_{\beta,n}\|^2]$ and the {bound on the} number of gradient steps to reach an ${\epsilon_g}$-stationary point is smaller than $T_{\mathrm{grad}}^\beta$ in Theorem~\ref{thm:stateful_warmstart}.
\end{theorem}
\begin{proof2}[Proof of Theorem~\ref{thm:element_memory}]
Under~\eqref{eq:element_memory} with episode-level updates, the steady-state memory is $\bar{m}_n = \mathbb{E}[\phi_n^*]$. The memory-augmented initialization error is presented in \eqref{eq:error_decomp}.
\begin{figure*}[!t]
\vspace{0.5em}
\begin{align}
e_{m,n}(t) &= \phi_n^*(t) - \beta_n\phi_n^*(t-1) - (1-\beta_n)u_n(t) - \gamma_n m_n(t-1) = e_{\beta,n}(t) - \gamma_n m_n(t-1)
\label{eq:error_decomp}
\end{align}
\vspace{-0.5em}
\end{figure*}
Taking variance and using the independence of $m_n(t-1)$ from the innovation $\boldsymbol{\mathcal{E}}(t)$, we obtain \eqref{eq:var_expand}.
\begin{figure*}[!t]
\vspace{-3em}
\begin{align}
\mathrm{Var}[e_{m,n}(t)] &= \mathrm{Var}[e_{\beta,n}(t)] + \gamma_n^2 \mathrm{Var}[m_n(t-1)] 
- 2\gamma_n\,\mathrm{Cov}[e_{\beta,n}(t),\, m_n(t-1)]
\label{eq:var_expand}
\end{align}
\vspace{-0.5em}
\hrule
\end{figure*}
$e_{\beta,n}(t) = \phi_n^*(t) - \phi_{n,\beta}^{(0)}(t)$ and $m_n(t-1)$ being a weighted average of past optimal phases lead to the error and historical mean having a negative correlation because of the Gauss--Markov persistence. This means when the current optimum is high, the historical mean is too, but the $\beta$-only initialization undershoots. So, $-2\gamma_n \mathrm{Cov}[e_{\beta,n}(t), m_n(t-1)] < 0$ holds true for sufficiently small $\gamma_n > 0$ values. 
By setting $\gamma_n < 2|\mathrm{Cov}[e_{\beta,n}, m_n]| / \mathrm{Var}[m_n]$, the variance reduction part takes over. This is substituted into the convergence proof from Theorem~\ref{thm:warmstart}.
\end{proof2}
Theorem~\ref{thm:element_memory} establishes that the distributed element memory $m_n(t)$ and the {surface inertia} coefficient $\beta$ are complementary, not redundant. {$\beta$ acts on the realized trajectory once per coherence interval, averaging the solver residual and smoothing the phase updates at the price of the factor $1/(1-\beta^2)$ in Theorem~\ref{thm:stateful_warmstart}}, while $m_n(t)$ further reduces the variance of the residual error by exploiting per-element cross-episode phase persistence at the episode timescale. Neither mechanism subsumes the other.
\begin{corollary}[Memory Phase Transition]
\label{cor:phase_transition}
Assuming A5 and A6, Theorem~\ref{thm:element_memory} shows that element memory minimizes the initialization error in case of correlation within the channels; however, there is no indication as to where this advantage switches sides. 
We identify a critical Doppler threshold as
\begin{equation}
{f_D^* = \frac{1}{2\pi T_s}J_0^{-1}\!\left(\rho_{\mathrm{th}}\right),\qquad \rho_{\mathrm{th}}=\sqrt{1 -
\frac{\gamma_n(1-\lambda)}{1+\gamma_n}},}
\end{equation}
\vspace{0.4em}
where $\mathrm{Var}[e_{m,n}(t)] < \mathrm{Var}[e_{\beta,n}(t)]$. This holds if $f_{D,\max} < f_D^*$, where $J_0^{-1}$ is the inverse of $J_0$ on $[0,j_{0,1}]$ and $j_{0,1}$ is the first zero of $J_0$; the expression is exact whenever $2\pi f_{D,\max}T_s\le j_{0,1}$. Past this threshold,
$m_n(t)$ will decorrelate a lot faster than it can be leveraged, and setting $\gamma_n = 0$ is strictly optimal.
\end{corollary}
The greater the value of $\lambda$, the greater $f_D^*$, expanding the Doppler range in which memory is beneficial, while an increase in $\gamma_n$ reduces it.
The meta-policy will learn $\gamma_n$ jointly with other surface parameters,
thereby self-regulating with respect to $f_D^*$ as expected based on the value of the learned $\beta^*$.
The proof of Corollary~\ref{cor:phase_transition}
is presented in Appendix~B.
\vspace{-6pt}
\subsection{Causal Intervention Filter}
\label{sec:cif}
We develop a rejection rule inspired by structural causal modeling (SCM) \cite{pearl2009causality} that tests the physical feasibility of the meta-policy before deploying its results on the surface. The meta-policy results are tested against physical consistency through three criteria: a power consistency criterion, a non-negativity condition, and correct phase boundaries. Any meta-policy phase result $\hat{\boldsymbol{\phi}}(t)$ is deemed invalid when violating any of the following:
\begin{subequations}\label{eq:feasibility}
\begin{alignat}{3}
&\|\mathbf{g}_{q,k}(\hat{\boldsymbol{\phi}}(t))\|_2^2
    &&\leq \Gamma_{q,k},
    &&\quad \forall k,
    \label{eq:feasibility_a} \\
&\mathcal{R}_k(\hat{\boldsymbol{\phi}}(t))
    &&\geq 0,
    &&\quad \forall k,
    \label{eq:feasibility_b} \\
&\hat{\phi}_n^\mathsf{r},\, \hat{\phi}_n^\mathsf{t}
    &&\in [0, 2\pi),
    &&\quad \forall n,
    \label{eq:feasibility_c}
\end{alignat}
\end{subequations}
where $\Gamma_{q,k} \triangleq N^2 Q\, {\zeta_{q}\zeta_{k}^\mathsf{r}}$ is the maximum physically realizable channel power implied by the large-scale coefficients.
Under the previous assumptions, the conditions are 
satisfied by a phase vector in $\mathcal{C}_{\mathrm{clip}}$ 
that nevertheless produces a physically invalid 
channel response, e.g., one exceeding the bound 
$\Gamma_{q,k}$. The causal filter rejects all 
such vectors.
\begin{proposition}
\label{prop:causal}
Let $\mathcal{C}_{\mathrm{clip}} = [0, 2\pi)^{2N}$ denote the set of phase vectors passing the phase-bound constraint alone, and let $\mathcal{C}_{\mathrm{causal}} \subseteq \mathcal{C}_{\mathrm{clip}}$ denote the subset additionally satisfying the power-anomaly and rate-phase constraints. Then $\mathcal{C}_{\mathrm{causal}} \subsetneq \mathcal{C}_{\mathrm{clip}}$.
There exist phase vectors in $\mathcal{C}_{\mathrm{clip}}$ that are rejected by the causal filter but accepted by clipping alone.
\end{proposition}
The proof of proposition~\ref{prop:causal} is presented in  Appendix C.
The proposition makes a conservative claim that the causal filter strictly enlarges the rejection set relative to clipping. It does not claim that every rejected vector is harmful nor that the filter eliminates all harmful vectors, only that it eliminates a strictly larger class than clipping alone.
\vspace{-0.6em}
\subsection{Meta-Learning Optimization}
The cross-episode learning formulation aims to learn the meta-policy parameters, considering three conflicting objectives: communication quality, efficiency of computation on hardware, and system stability. Here $\mathbf{s}_{\mathrm{new}}$ is the context vector that characterizes the current environment condition and is not an optimization variable. 
The meta objective function is defined by
\begin{align}
\max_{\theta}\; \mathcal{J}_{\mathrm{meta}}(\theta, \mathbf{s}_{\mathrm{new}}) &= \mathcal{J}_{\mathrm{rate}} - \mathcal{J}_{\mathrm{energy}} - \mathcal{J}_{\mathrm{penalty}}, \nonumber\\[-0.4em]
\mathcal{J}_{\mathrm{rate}} &= \omega_{1} \sum_{k=1}^K \log_2(1 + \mathrm{SINR}_k(\hat{\mathbf{a}})), \nonumber\\[-0.4em]
\mathcal{J}_{\mathrm{energy}} &= \omega_{2} \sum_{n=1}^N c_n\!\left[(\phi_n^{\mathsf{r}}(t))^2 + (\phi_n^{\mathsf{t}}(t))^2\right], \nonumber\\[-0.4em]
\mathcal{J}_{\mathrm{penalty}} &= \lambda_{\mathrm{ris}}\sum_{n=1}^N P_n(\alpha_n(t)) \nonumber\\[-0.4em]
&\quad + \lambda_c\|\boldsymbol{u} - \boldsymbol{\phi}(t-1)\|_2^2,
\label{eq:meta_obj}
\end{align}
where the causal filter rejects any $\hat{\mathbf{a}}$ that violates the feasibility conditions before this objective is analyzed.
The meta-policy parameters parameterize a multilayer perceptron (MLP) with ReLU activations, which maps the concatenated context vector 
$C \in \mathbb{R}^{D_C}$ to the phase action 
$\hat{\mathbf{a}} \in \mathbb{R}^{2N}$, where 
$D_C = D_s + D_t + 2N$ concatenates the scenario 
features, compressed tag vector, and current 
phase state.
The joint optimization is defined as
\begin{subequations}
\begin{align}
\boldsymbol{\phi}^*(t) &=
\argmax_{\boldsymbol{\phi},\,\boldsymbol{\alpha}}
F(\boldsymbol{\phi},\boldsymbol{\alpha}),
\label{eq:joint_phi}\\[4pt]
\theta^* &=
\argmax_{\theta}\;
\mathcal{J}_{\mathrm{meta}}
(\theta,\mathbf{s}_{\mathrm{new}}),
\label{eq:joint_theta}
\end{align}
\end{subequations}
where~\eqref{eq:joint_phi} is solved {in every coherence interval by warm-started projected gradient ascent, which returns a stationary point in the sense of A3 and not necessarily the global maximizer; $\boldsymbol{\phi}^*(t)$ denotes this stationary point throughout the paper}, and \eqref{eq:joint_theta} is solved online via gradient ascent.
The first line solves the intra-episode phase optimization, while the second solves the 
cross-episode meta-policy update.
\vspace{-0.7em}
\subsection{Online Mirror-Descent Tag-Weight Learning}
\label{sec:omd}
Using fixed importance weights $\mathcal{W}_j$ to represent the novelty score is problematic as the relative importance of various tag dimensions varies for deployment scenarios. In addition, there is no mechanism to improve weights over time. We replace this with online mirror descent over the probability simplex, which provides adaptive weights with a provable regret bound.
This loss encourages higher weights for tag dimensions that improve retrieval similarity and downstream sum-rate performance. The instantaneous novelty loss is represented as
\vspace{-0.6em}
\begin{equation}
\ell_t(\mathcal{W}) = -\sum_{j=1}^{D_t} \mathcal{W}_j \cdot 
\tilde{\mathbf{t}}_{\mathrm{new}}^{(t)}[j]\cdot
\tilde{\mathbf{t}}_{i^*}[j] \cdot \nu_{t},
\end{equation}
where $\mathcal{W}_j \in \Delta^{D_t}$ is the importance weights maintained and updated by the online mirror-descent procedure, the product $\tilde{\mathbf{t}}_{\mathrm{new}}^{(t)}[j]\cdot\tilde{\mathbf{t}}_{i^*}[j] \in \{0,1\}$ measures bitwise tag agreement, $\tilde{\mathbf{t}}_{i^*}$ is the compressed tag vector, $\tilde{\mathbf{t}}_{\mathrm{new}}^{(t)}$
 is the compressed tag vector of the current episode at episode t, and $\nu_{t} = \mathcal{R}_{\mathrm{sum}}(t)$ denotes the sum-rate reward achieved at episode $t$.
The mirror-descent update gives
\begin{equation}
\mathcal{W}_j^{(t+1)} = \frac{\mathcal{W}_j^{(t)} \exp(-\eta_{\mathcal{W}} \nabla_j \ell_t(\mathcal{W}))}{\sum_{j'} \mathcal{W}_{j'}^{(t)} \exp(-\eta_{\mathcal{W}} \nabla_{j'} \ell_t(\mathcal{W}))},
\label{eq:mirror_descent}
\end{equation}
where $\eta_{\mathcal{W}}$ is the 
step size and $\nabla_j \ell_t(\mathcal{W})$ 
is the gradient of $\ell_t(\mathcal{W})$ 
with respect to $\mathcal{W}_j$.
\begin{theorem}
\label{thm:regret}
Under A8, if $L = \max_{t,\mathcal{W}} 
\|\nabla_{\mathcal{W}} \ell_t(\mathcal{W})\|_2$ 
is the Lipschitz constant of the loss function. With a training episode budget $T$ and step size $\eta_{\mathcal{W}} = 
\sqrt{2\ln D_t / (TL^2)}$, then the accumulated regret bound holds for 
all $t \leq T$ as follows:
\begin{equation}
\sum_{t=1}^T \ell_t(\mathcal{W}^{(t)}) - 
\min_{\mathcal{W} \in \Delta^{D_t}} 
\sum_{t=1}^T \ell_t(\mathcal{W}) \leq 
L\sqrt{2T\ln D_t}.
\label{eq:regret_bound}
\end{equation}
\end{theorem}
\begin{proof2}[Proof of Theorem~\ref{thm:regret}]
This follows from the standard mirror-descent 
regret analysis with negative entropy as the 
mirror mapping on $\Delta^{D_t}$~\cite{shalev2025online}, 
where the Bregman divergence satisfies 
$D_{\mathrm{KL}}(\mathcal{W} \| \mathcal{W}^{(1)}) 
\leq \ln D_t$ for any $\mathcal{W} \in \Delta^{D_t}$ 
under uniform initialization.
\end{proof2}
The regret bound $L\sqrt{2T\ln D_t}$ grows as 
$\mathcal{O}(\sqrt{T})$, meaning the per-episode regret $\to 0$ as $T\to\infty$, so the weights asymptotically match the best fixed weights in 
hindsight. 
A gradient step forms the effective channels of the $K$ users at the serving AP, at cost $\mathcal{O}(KQN)$, and the pairwise interference terms, at cost $\mathcal{O}(K^2Q)$. Considering the tag retrieval over the $T_E$ stored episodes, the cost per episode is expressed as$\mathcal{O}\big(T_{\mathrm{grad}}KQ(N+K)+T_ED_t\big)$. Assuming $M$ jointly processed APs, the first term is multiplied by $M$. Consequently, the time complexity of our algorithm per episode is $\mathcal{O}(T_{\mathrm{grad}} \cdot KN)$ in terms of elements and users for $N \gg K$,
where $T_{\mathrm{grad}} = \mathcal{O}(\beta_F^2(1-\rho_c^2)C_\Delta^2/\epsilon_g^2)$
as given in Theorem~\ref{thm:warmstart}, against $\mathcal{O}(\beta_F^2\|\boldsymbol{\phi}_{\max}\|_2^2/\epsilon_g^2 \cdot KN)$
for cold-start MAML, which represents a reduction of the bound by a factor of
$\|\boldsymbol{\phi}_{\max}\|_2^2/((1-\rho_c^2)C_\Delta^2)$
that grows with $\rho_c$. The memory footprint of $\mathcal{M}$
is $\mathcal{O}(T_E(D_s + D_t + 2N))$, linear in $N$ and
independent of $K$, plus $N$ scalars for the element memory. Tag retrieval, the element-memory update, and the update of $\beta$ do not depend on $K$, and only the last two grow with $N$ linearly. The dependence on $N$ can be seen in Table~\ref{tab:gradsteps}, where the number of gradient steps remains almost constant as $N$ varies from 64 to 256, whereas the runtime increases. In this set of simulations, $K$ is kept constant, hence its dependence comes from the cost equation.

Learning from experience occurs through three complementary mechanisms operating at different timescales, which are (i)~within each episode, projected gradient ascent~\cite{nesterov2004introductory} warm-started from $\mathcal{M}$ via~\eqref{eq:warmstart} refines the phase configuration; (ii)~across episodes, element memory states $\{m_n(t)\}$ via~\eqref{eq:element_memory} tighten the warm-start gap as shown in Theorem~\ref{thm:element_memory}; and (iii)~over long horizons, the meta-policy updates $\theta$, $\beta$, $\beta_n$, $\gamma_n$, $\lambda$, and tag weights $\mathcal{W}$ via~\eqref{eq:mirror_descent} from accumulated experience in $\mathcal{M}$.
These three mechanisms form nested loops at varying speeds: each inner loop converges before the outer loop updates, where better tag weights reduce retrieval distance, reducing $\Delta_{\rho_c}$ and the gradient steps needed, so that the loops reinforce one another.
The details of our algorithm are explained in Algorithm \ref{meta2l}.
\begin{algorithm}
\caption{Meta2L for Stateful STAR-RIS Optimization}
\label{meta2l}
\begin{algorithmic}[1]
\footnotesize
\Require $\mathcal{M}$, $\beta$, $\lambda$, $\gamma_n$,
         $\mathcal{W} \gets \mathbf{1}/D_t$, $m_n(0) \gets 0$
\Ensure $\boldsymbol{\phi}^*(t)$ for each coherence interval $t$
\For{$t = 1, 2, \dots$}
    \State Observe $\mathbf{H}_{{q}}(t)$; compute $\rho_c$; extract $\tilde{\mathbf{t}}_{\mathrm{new}}$
    \State Retrieve $i^*, i^{**}$ from $\mathcal{M}$ via $\mathcal{W}$
   \State $\boldsymbol{u}(t) \gets \rho_c 
\boldsymbol{\phi}^*_{i^*} + (1-\rho_c)
\boldsymbol{\phi}^*_{i^{**}}$ 
\Comment{warm-start command~\eqref{eq:warmstart}}
\State $\phi_n^{(0)} \gets \beta_n\phi_n^*(t-1) + (1-\beta_n)u_n(t) + \gamma_n m_n(t-1)$ 
\Comment{memory-augmented init~\eqref{eq:memory_phase}}
    \If{(\ref{eq:feasibility}) violated}
        \State $\boldsymbol{\phi}^{(0)}(t) \gets$ random initialization
    \EndIf
    \State $\boldsymbol{\phi}^*(t) \gets \mathrm{clip}(\boldsymbol{\phi}^{(0)}(t) + \eta_{\phi}\nabla_{\boldsymbol{\phi}} F,\, 0,\, 2\pi)$ until $\|\nabla F\|<\epsilon$ \Comment{\eqref{eq:boundary_sol}}
        \State $\boldsymbol{\phi}(t) \gets \beta\boldsymbol{\phi}(t-1) + (1-\beta)\boldsymbol{\phi}^*(t)$ \Comment{\eqref{eq:stateful_ris}}
    \State Update $\beta$ with the gradient (\ref{eq:beta_update})
    \State $m_n(t) \gets \lambda m_n(t-1) + (1-\lambda)\phi_n^*(t),\ \forall n$ \Comment{\eqref{eq:element_memory}}
    \State Update $\theta$ via $\nabla_\theta \mathcal{J}_{\mathrm{meta}}$; store in $\mathcal{M}$
    \State $\mathcal{W}_j^{(t+1)} \gets \mathcal{W}_j^{(t)} e^{-\eta_{\mathcal{W}} \nabla_j \ell_t} / \sum_{j'} \mathcal{W}_{j'}^{(t)} e^{-\eta_{\mathcal{W}} \nabla_{j'} \ell_t}$ \Comment{\eqref{eq:mirror_descent}}
\EndFor
\end{algorithmic}
\end{algorithm}
\vspace{-0.5em}
\section{Results and Performance Evaluation}
\label{sec5}
\vspace{-0.1em}
\begin{table}[t!]
\centering
\caption{Simulation Parameters}
\label{params}
\scriptsize
\renewcommand{\arraystretch}{1.0}
\begin{tabular}{ll}
\toprule
\textbf{Parameter} & \textbf{Value} \\
\midrule
\multicolumn{2}{l}{\textit{Communication System}} \\
APs $|$ UEs ($M|K$) & 4 $|$ 32 \\
Antennas per AP ($Q$) & 64 (ULA) \\
RIS elements ($N$) & 16, 64, 100, 256 \\
Frequency $|$ Bandwidth & 28 GHz $|$ 10 MHz \\
TX Power $|$ Noise & 10 dBm $|$ $-90$ dBm \\
Velocity $|$ Rician $K$ & 0--30 m/s $|$ 3 dB \\
\midrule
\multicolumn{2}{l}{\textit{Meta2L Configuration}} \\
Memory ($T_E$) $|$ L2 size & 1000 $|$ 50 \\
Tag dimension ($D_t$) & 24 \\
Hidden layers & $2 \times 256$ (ReLU) \\
Gradient step size $\eta_{\phi}$ & $0.01$ \\
Thresholds ($\theta_h$, $\theta_l$) & 0.75 $|$ 0.35 \\
Step size $\eta_{\mathcal{W}}$ & $\sqrt{2\ln 24/(TL^2)}$ \\
Weights $\omega_1, \omega_2$ & 10.0 $|$ 1.0 \\
Surface memory $\beta$ & 0.0, 0.3, 0.6, 0.9 \\
Overhead weight $\lambda_c$ & 0.1 \\
$\alpha_{\mathrm{act}}$ $|$ $\lambda_{\mathrm{ris}}$ & 2.0 $|$ 0.5 \\
{Mode in simulations} & {unit gain} \\
{Energy splitting $\eta_n$} & {0.5} \\
{Update interval $T_s$} & {0.1 ms} \\
\midrule
\multicolumn{2}{l}{\textit{Training \& Evaluation}} \\
Episodes (train $|$ test) & 800 $|$ 200 \\
Monte Carlo trials & 10 \\
Mobility regimes & Ped., Mixed, Veh. \\
\midrule
\multicolumn{2}{l}{\textit{DRL-PPO Baseline}} \\
Episodes & 2000 ({$2.5\times$} Meta2L) \\
Network $|$ Optimizer & $3\times256$ ReLU $|$ Adam ($3\!\times\!10^{-4}$) \\
$\gamma_{\mathrm{PPO}}$ $|$ $\lambda_{\mathrm{GAE}}$ $|$ $\epsilon$ & 0.99 $|$ 0.95 $|$ 0.2 \\
Replay buffer & 10,000 \\
\bottomrule
\end{tabular}
\end{table}
The simulator runs on Python and makes use of NumPy, PyTorch, Scikit-learn, and SciPy. We average the results over ten Monte Carlo trials and show error bars to indicate $\pm$ one standard deviation. Table \ref{params} provides the full details of the simulation parameters.
While generating the scenario, in each trial, the LoS angles of arrival and departure are drawn uniformly from $[-60^\circ,60^\circ]$ for every link and then kept fixed throughout the trial. The NLoS components evolve according to (\ref{eq:gauss_markov}), with $T_s=0.1$~ms and the RIS-to-AP correlation determined by the mean UE velocity. For each UE and trial, the velocity is drawn from $[0.5,5]$~m/s for the pedestrian scenario, $[20,30]$~m/s for the vehicular scenario, or $[0.5,30]$~m/s for the mixed scenario. The episode tag vector is constructed from thresholded velocity, channel power, array, network, and quality-of-service attributes. Each trial uses its own fixed random seed. The simulation code and the scripts used to generate the figures and tables are available in \cite{hassan2026code}.
Our benchmarks include SIO, APG, AO under robust beamforming, GMLB, and DRL-PPO. SIO and APG correspond to classic optimization methods with no learning ability; AO corresponds to robust optimization without any cross-episode learning; GMLB corresponds to gradient-based meta-learning without Doppler warm-up and regret-bounded weights; DRL-PPO corresponds to the state-of-the-art learning scheme without cross-episode learning; and random Phase corresponds to the lower bound. DRL-PPO is trained over 2000 episodes, which is 2.5 times that of Meta2L's 800 training episodes, and evaluated at the end of the DRL-PPO training phase as opposed to at the point where Meta2L converges, thus no disadvantage from having a bigger budget. $\lambda_c$ is set small to allow fast adaptation while penalizing large innovations, and $\lambda_{\mathrm{ris}}$ balances hardware energy against beamforming gain. The four values of $\beta$ occupy the full inertia range from the memoryless case ($\beta=0$) to near-static ($\beta=0.9$).
In order to verify the results obtained by simulation,
we have performed two hardware-aware evaluations. These include a latency estimate and a simulation of circuit impairments. The Meta2L algorithm and all baselines are implemented and evaluated in simulation. The FPGA part is an analytical mapping of the critical-path operations onto the resources of the device and does not involve synthesis or measurement. The varactor quantization and settling are simulated with the model of A7. 
The first test involves mapping all operations along
the critical path of the Meta2L to the Xilinx Zynq UltraScale$+$ MPSoC (device XCZU9EG)~\cite{xilinx2022zynq, xilinx2023zynq}, wherein its configurable-logic fabric
offers the possibility of utilizing {DSP48E2} slices with the capacity to execute one multiplication/accumulation of ${27{\times}18}$ bits per clock cycle~\cite{xilinx2018ultrascale}; with a fabric clock of $200$ MHz, every single such operation takes exactly $5$~ns, whereas cycle numbers per stage follow directly from the arithmetic operations in the algorithm. This first test is an analytical cycle-count estimate.
The second experiment consists of two varactor circuits' impairments considered for the stateful controller as well as for the memoryless baseline.
Every continuous phase is rounded to the nearest of the $2^3$ levels of a $3$-bit quantizer, and the RC transient of the biasing circuit is modeled as a convex combination of the previous and target phase states.
\begin{figure}[t!]
    \centering
    \includegraphics[width=\columnwidth]{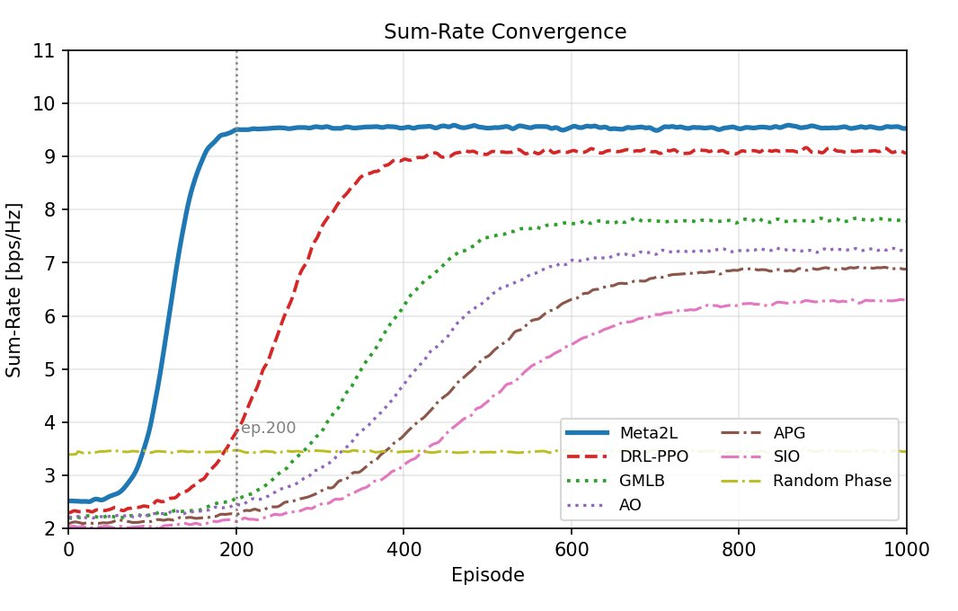}
    \caption{Average sum-rate (bps/Hz) versus episode count 
under mixed mobility, compared with baselines. Error bars indicate $\pm 1$ standard deviation over 
10 Monte Carlo trials.
}
    \label{sum}
\end{figure}
Fig.~\ref{sum} displays the average sum-rate vs.\ episode count, showing that Meta2L converges to $\approx 9.5$~bps/Hz by episode 200 while DRL-PPO converges more slowly ($\approx 8.1$~bps/Hz) and GMLB plateaus at $\approx 7.8$~bps/Hz. The other baselines fall well below these values.
\begin{figure}[t!]
    \centering
        \includegraphics[width=\columnwidth, height = 2.2in]{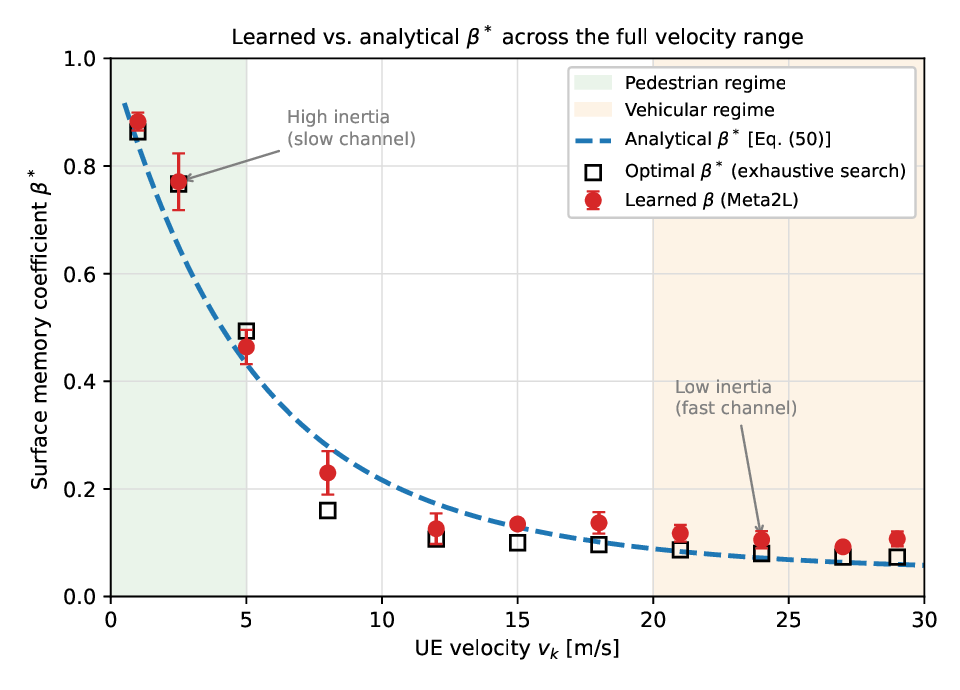}
    \caption{Learned surface inertia coefficient $\beta$ versus UE
velocity, compared against the analytical value from~\eqref{eq:beta_opt}. The optimum was found by exhaustive search over $\beta$ on the same channel realizations (mean $\pm$ one standard deviation over three trials).}

    \label{fig5}
\end{figure}
In Fig.~\ref{fig5}, $\beta$ is learned online using (\ref{eq:beta_update}) in a single pass through each channel sequence. The comparison is done against two benchmarks: the closed-form solution from (\ref{eq:beta_opt}) with the sole constant being $C_\Delta^2/\sigma_w^2 = 17.6$, which is estimated using least squares based on the search results, and the minimum of the mean $J_t(\beta)$ on a $\beta$ grid consisting of 100 points. The learned values are within 0.03 of the optimum on average, without using $\beta_F$, the channel statistics, or any future channel samples.
For pedestrian channels ($v_k\le 2.5$~m/s, $\rho_c\ge 0.995$), the method learns high inertia, with $\beta\approx 0.77$--$0.88$. For vehicular channels ($v_k\ge 21$~m/s, $\rho_c=0.40$--$0.66$), it instead learns much lower inertia, with $\beta\approx 0.09$--$0.12$. This trend agrees with the expected behavior: slower channel variations allow the surface to retain its previous state for longer. The transition occurs between 5 and 8~m/s. At 28~GHz with $T_s=0.1$~ms, the channel correlation has already decreased to $\rho_c=0.95$ at 8~m/s.
\begin{figure}[t!]
    \centering
    \includegraphics[width=\columnwidth, height=2.3in, 
    keepaspectratio=false]{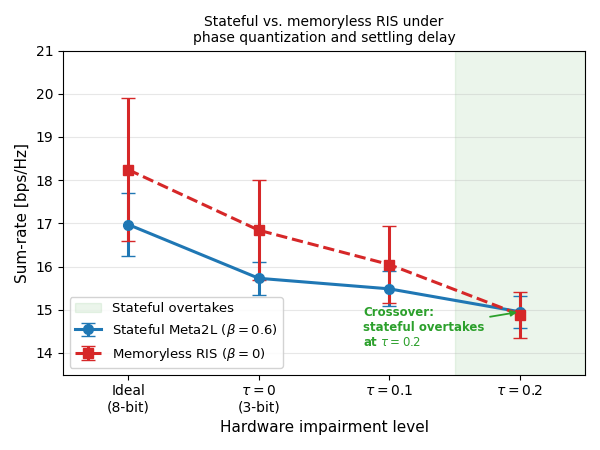}
    \caption{Sum rate of stateful RIS at $\beta{=}0.6$ vs.\ memoryless RIS under progressive hardware impairments}
    \label{sens}
\end{figure}
Under ideal hardware, the memoryless system in Fig. ~\ref{sens} performs 
better since unconstrained phase jumps  exploit the instantaneous channel. As settling 
delay increases, the gap narrows and the stateful counterpart overtakes 
at $\tau{=}0.2$ ($+0.4\%$), since 
$\beta$ anticipates partial settling by design 
while the memoryless controller optimizes for 
an ideal jump that never fully realizes (see also Fig.~\ref{fig66}).
\begin{figure}[t!]
    \centering
    \includegraphics[width=\columnwidth, height=2.2in, 
    keepaspectratio=false]{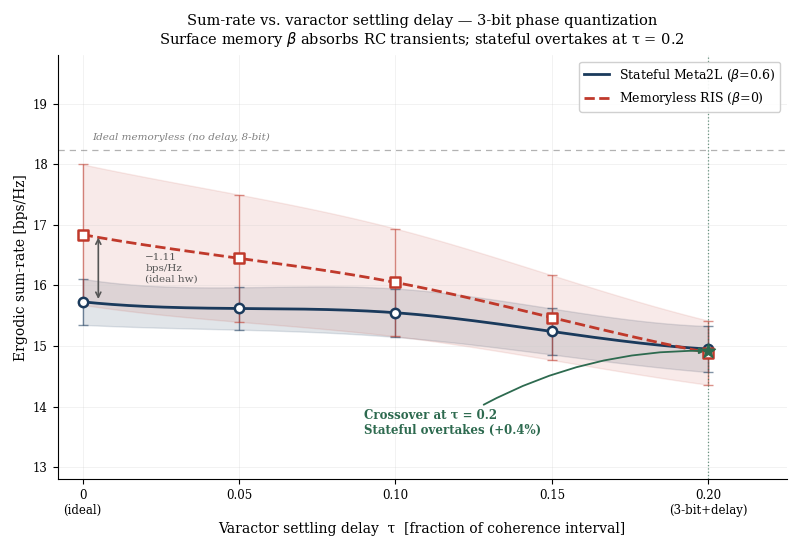}
    \caption{Ergodic sum-rate as a function of varactor settling delay for Stateful Meta2L and the memoryless RIS baseline under 3-bit phase quantization.}
    \label{fig66}
\end{figure}
Fig. \ref{fig66} shows that the classical memoryless model leads by 1.11 bps/Hz for the ideal hardware case due to unconstrained phase jumps. As the settling delay increases, the memoryless model's performance significantly degrades {because it optimizes for phase transitions that the varactor RC transients do not realize}. The surface {inertia} coefficient smooths the phase trajectory to anticipate partial settling, which causes the gap to close monotonically until the stateful model overtakes.

Our model has the settling time $\tau$ defined as the proportion of the previous phase remaining at the end of the update period $T_s$. For a first-order bias circuit with time constant $\tau_{\mathrm{RC}}$, this fraction is $\tau=e^{-T_s/\tau_{\mathrm{RC}}}$. Therefore, for $T_s=0.1$ ms, $\tau=0.1$, $\tau=0.2$ represent $\tau_{\mathrm{RC}}=43 ~\mu$s and $62 ~\mu$s, respectively. In this case, $\tau$ represents the settling of an individual component to the new biasing instruction, which includes the reaction of the tuner, its bias network, and the driver. It does not include the time taken to program the whole surface via the control interface, which is considered independently as control latency.
We did not find reported settling times for STAR-RIS prototypes. In varactor-based RIS, the phase is set by a DC bias applied through a filtered bias network, and measurements of RIS with embedded varactors characterize their nonlinear behavior, such as third-order intermodulation \cite{manna2026varactor}. The varactor itself responds much faster than $T_s$, so settling is set mainly by the bias network.
Therefore, the $\tau\le0.2$ range considered in Figs.~\ref{sens} and~\ref{fig66} corresponds to bias networks with time constants below about 62~$\mu$s. If the hardware settles more slowly, $T_s$ must be increased accordingly, which also reduces $\rho_c$ for a given user speed. The benefit that the stateful model has compared to the memoryless model should be considered in relation to the settling mode that the hardware can support.
\begin{figure}[t!]
    \centering
        \includegraphics[width=\columnwidth, height=2.2in,
    keepaspectratio=false]{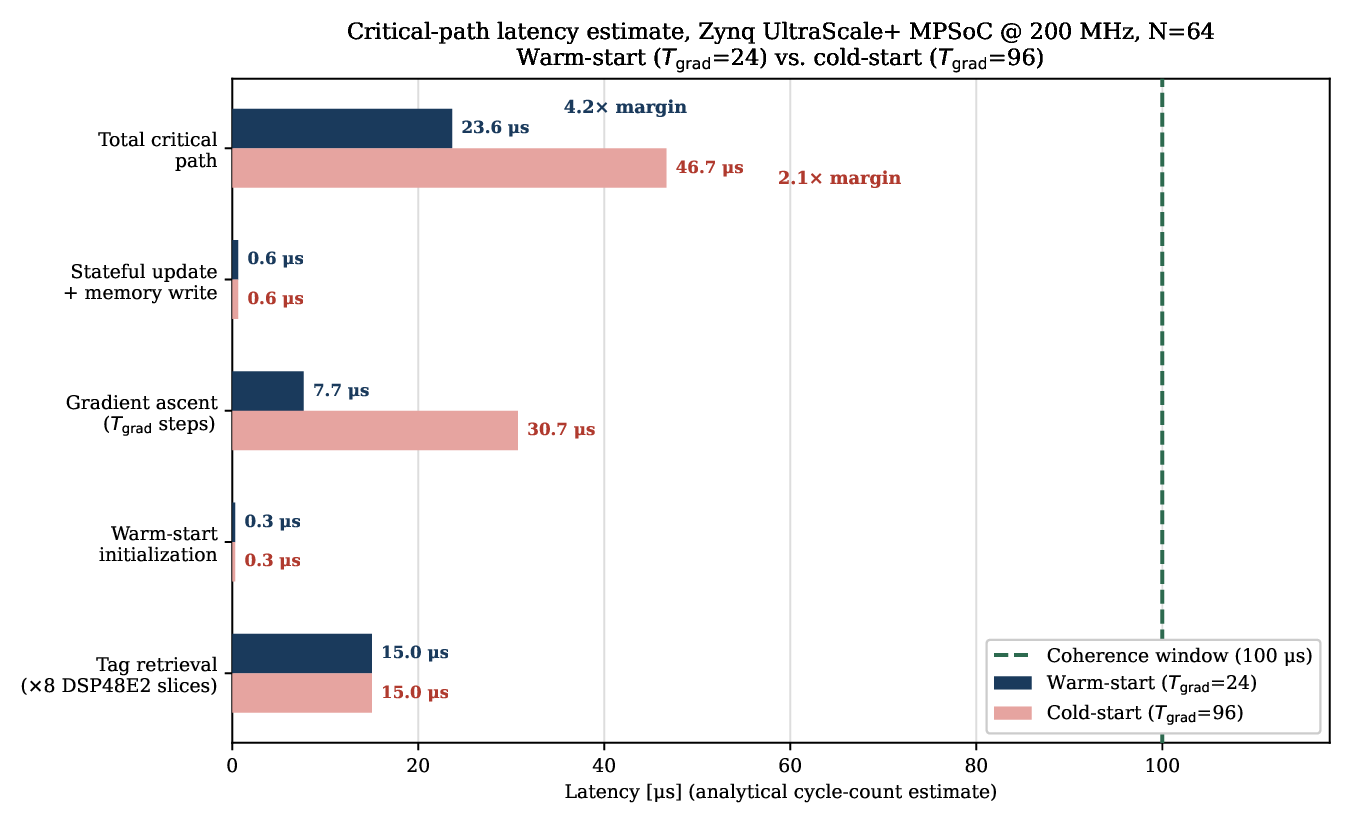}
    \caption{Critical path latency breakdown of Meta2L on the Xilinx Zynq UltraScale+ MPSoC (analytical cycle-count estimate). Each operation is mapped to {DSP48E2} slices and execution time is shown for both the warm-start path and cold-start path.}
    \label{mirror}
\end{figure}
Fig. \ref{mirror} demonstrates that tag retrieval dominates both paths at 15.0 $\upmu$s due to the parallel execution of {DSP48E2}. The gradient ascent improves latency, which drops from 30.7 $\upmu$s to {7.7} $\upmu$s under warm-start, which represents {a $4\times$} reduction {(96 versus 24 gradient steps, as in Table~\ref{tab:gradsteps}). This estimate indicates that the computational critical path is compatible with the coherence window, and it} brings the total critical path to {23.6} $\upmu$s, which represents a {$4.2\times$} safety margin within the coherence window.
These estimates have four primary shortcomings. Firstly, it takes into account only arithmetic cycles, the assumption being that all data stays on-chip, and each gradient step takes an $N$-way parallel datapath, meaning $N$ cycles per step. If there are fewer parallel multipliers, then it will take longer to perform each step. Secondly, this estimation does not take into consideration the CSI acquisition and the delay of the AP-to-surface control link. Thirdly, the comparison is done solely in terms of the channel coherence window. Lastly, these are analytical estimations and not synthesis or measurements. An actual implementation and synthesis on the target device are left for future work.
\begin{figure}[t!]
    \centering
    \includegraphics[width=\columnwidth]{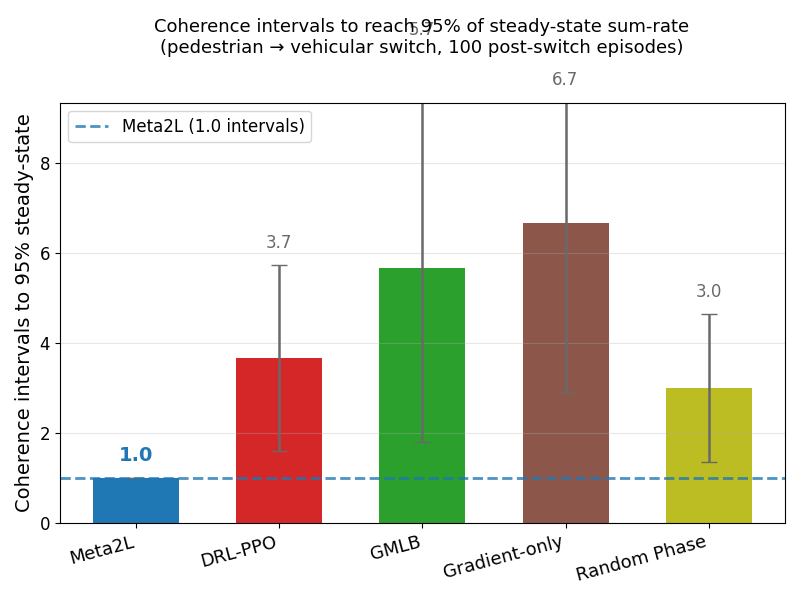}
    \caption{Evolution of sum-rate following mobility-induced channel changes, demonstrating how Meta2L achieves steady-state performance quicker compared to baselines.}
    \label{fig4}
\end{figure}
Fig.~\ref{fig4} shows the number of coherence intervals required to reach 95\% of steady-state sum-rate following a mobility-induced channel change, showing that Meta2L adapts in $\approx 1$ interval due 
to its episodic memory buffer; all baselines need $3$--$7$ intervals.

Table~\ref{tab:ablation} illustrates the results from our ablation study, which includes energy efficiency (EE) {and} adaptation speed (Ad.). Our full system outperforms the other configurations by attaining the highest SR, lowest EE cost, and fastest adaptation at once, indicating the effectiveness of each component independently and jointly. The Ad.\ column is the number of coherence intervals required to reach 95
The ablation rows are arranged in pairs, where rows (2) and (6) correspond to episodic retrieval, rows (3) and (7) to the Doppler warm start, rows (4) and (8) to mirror-descent tag weighting, and rows (5) and (9) to the causal filter. The first row in each pair adds the component to the other components, while the second removes it. This lets us compare the results with and without the same component.
We consider the two sources of surface memory separately. The effect of surface inertia is shown by the learned $\beta$ in Fig.~\ref{fig5} and by the stateful and memoryless cases in Figs.~\ref{sens} and~\ref{fig66}. The memory of the individual elements is considered in Theorem~\ref{thm:element_memory} and Corollary~\ref{cor:phase_transition}, and its decay parameter $\lambda$ is varied in Table~\ref{tab:gradsteps}.
\begin{table}[!htbp]
\centering
\caption{Ablation Study: Incremental Build-Up and Leave-One-Out
(Mean $\pm$ Std, 200 Decision Episodes, 10 Trials, Mixed Mobility){. Each build-up row adds one component to the row above; each leave-one-out row removes one component from the full system.}}
\label{tab:ablation}
\scriptsize
\renewcommand{\arraystretch}{1.1}
\setlength{\tabcolsep}{3pt}
\begin{tabular}{|l|c|c|c|}
\hline
\rowcolor{gray!25}
\textbf{Configuration} & \textbf{SR} & \textbf{EE} &
\textbf{Ad.} \\
\rowcolor{gray!25}
& \textbf{[bps/Hz]} & \textbf{[n.]} &
\textbf{[i.]} \\
\hline
\multicolumn{4}{|l|}{\textit{Incremental build-up}} \\
\hline
(1) No meta-learning
    & $5.9\pm1.3$ & $1.09$ & $21.0$ \\
(2) + Episodic memory
    & $7.0\pm1.0$ & $1.06$ & $14.5$ \\
(3) + Warm-starting
    & $8.2\pm0.9$ & $1.04$ & $7.1$ \\
(4) + Mirror-descent
    & $8.7\pm0.6$ & $1.02$ & $5.3$ \\
(5) + Causal = \textbf{Full}
    & $\mathbf{9.5\pm0.3}$ & $\mathbf{1.00}$
        & {$\mathbf{1.0}$} \\
\hline
\multicolumn{4}{|l|}{\textit{Leave-one-out}} \\
\hline
(6) w/o memory (MAML)
    & $7.8\pm1.1$ & $1.04$ & $12.1$ \\
(7) w/o warm-start
    & $8.2\pm1.0$ & $1.02$ & $9.6$ \\
(8) w/o mirror-descent
    & $8.9\pm0.6$ & $1.01$ & $4.0$ \\
(9) w/o causal filter
    & $9.2\pm0.8$ & $1.00$ & $3.3$ \\
\hline
\end{tabular}
\end{table}
In Table~\ref{tab:gradsteps}, the mean number of steps 
required by projected gradient ascent to reach a stationary 
point of $\epsilon=0.01$ is provided based on empirical 
results from 200 iterations. Meta2L needs fewer gradient 
ascents than MAML and GMLB, resulting in a $4\times$ 
decrease in gradient steps compared to cold-start MAML and 
a $2.3\times$ decrease compared to GMLB, which is in line 
with the warm-start improvement proved in 
Theorem~\ref{thm:warmstart}. From the scalability analysis 
results, we can see that the number of steps remains constant 
for all array dimensions tested, while runtime increases 
proportionally to $N$, and sum-rate increases monotonically. Regarding sensitivity 
analysis, we found that runtime is insensitive to the value 
of $\lambda$, while non-monotonic sum trend validates the phase transition threshold derived in 
Corollary~\ref{cor:phase_transition}.
For the evaluation of physical realizability under realistic 
circuit-level constraints, we simulate $3$-bit varactor phase quantization 
and settling delay $\tau \in \{0.0, 0.1, 0.2\}$ using a convex 
combination of the current and future phase states of the phase shifters,
which models their RC transient dynamics. Both effects are used uniformly 
on both the stateful Meta2L structure and the memoryless cold start baseline 
as depicted in the lower portion of Table~\ref{tab:gradsteps}. 
The stateful--memoryless difference is reduced systematically from $-7.0\%$ to $+0.4\%$ at $\tau{=}0.2$, 
confirming that the surface inertia coefficient $\beta$ 
provides implicit robustness against the settling effects.
In contrast, the memoryless approach maximizes the ideal instantaneous 
jump in phase that is never achieved due to delay.
\begin{table}[!htbp]
\centering
\caption{Gradient Steps, Runtime, Sensitivity,
and Hardware Validity Analysis
(Mean $\pm$ Std, 200 Decision Episodes,
Mixed Mobility). Hardware validity section
columns report Stateful SR $|$ Memoryless SR
$|$ Gain (\%) under phase quantization and
settling delay $\tau$.}
\label{tab:gradsteps}
\scriptsize
\renewcommand{\arraystretch}{1.1}
\setlength{\tabcolsep}{4pt}
\begin{tabular}{|l|c|c|c|}
\hline
\rowcolor{gray!25}
\textbf{Method / Configuration} & \textbf{Grad.\ Steps} &
\textbf{RT [ms]} & \textbf{SR [bps/Hz]} \\
\hline
\multicolumn{4}{|l|}{\textit{Complexity comparison}} \\
\hline
Meta2L (Full)
    & $24 \pm 4$  & $23.0 \pm 1.4$ & $17.8 \pm 2.2$ \\
w/o warm-start (MAML)
    & $96 \pm 18$ & $14.4 \pm 2.7$ & $7.8 \pm 1.1$ \\
GMLB
    & $55 \pm 11$ & $8.3 \pm 1.6$  & $7.8 \pm 0.9$ \\
AO
    & $62 \pm 9$  & $9.2 \pm 1.4$  & $7.1 \pm 1.0$ \\
DRL-PPO
    & $\text{---}^{\dagger}$ & $38.0 \pm 4.1$ & $8.1 \pm 0.7$ \\
\hline
\multicolumn{4}{|l|}{\textit{Scalability: varying $N$}} \\
\hline
$N{=}64$
    & $24 \pm 4$ & $23.0 \pm 1.4$ & $1.00\times$ \\
$N{=}100$
    & $26 \pm 4$ & $23.6 \pm 1.9$ & $1.04\times$ \\
$N{=}128$
    & $27 \pm 3$ & $26.9 \pm 1.0$ & $1.06\times$ \\
$N{=}256$
    & $27 \pm 5$ & $38.8 \pm 1.3$ & $1.13\times$ \\
\hline
\multicolumn{4}{|l|}{\textit{Sensitivity: varying $\lambda$}} \\
\hline
$\lambda{=}0.10$
    & $19 \pm 4$  & $23.7 \pm 2.0$ & $16.35 \pm 1.35$ \\
$\lambda{=}0.30$
    & $27 \pm 9$  & $25.6 \pm 1.0$ & $17.84 \pm 2.15$ \\
$\lambda{=}0.70$
    & $24 \pm 5$  & $26.3 \pm 1.5$ & $15.66 \pm 0.58$ \\
$\lambda{=}0.95$
    & $30 \pm 11$ & $24.3 \pm 0.6$ & $17.39 \pm 2.53$ \\
\hline
\multicolumn{4}{|l|}{\textit{Hardware validity:
Stateful Meta2L vs.\ Memoryless RIS}} \\
\hline
Ideal (8-bit, no delay)
    & $16.97\pm0.73$ & $18.24\pm1.65$ & $-7.0\%$ \\
3-bit quantization
    & $15.73\pm0.38$ & $16.84\pm1.16$ & $-6.6\%$ \\
3-bit $+$ 10\% delay
    & $15.48\pm0.41$ & $16.05\pm0.89$ & $-3.5\%$ \\
3-bit $+$ 20\% delay
    & $\mathbf{14.95\pm0.38}$ & $14.89\pm0.53$
    & $\mathbf{+0.4\%}$ \\
\hline
\end{tabular}
\\[3pt]
\raggedright\scriptsize
$^{\dagger}$DRL-PPO does not use projected
gradient ascent; steps not applicable.
Gradient steps reflect $N{=}64$, $K{=}32$.
Hardware validity columns:
Stateful SR $|$ Memoryless SR $|$ Gain (\%).
\end{table}
\begin{table}[!htbp]
\centering
\caption{Meta2L Hardware Latency Estimation
(Zynq UltraScale+, 200~MHz, $D_t{=}24$, $T_E{=}1000$). Warm-start cuts gradient steps from $96$ to ${24}$, {giving a $4.2\times$ margin against the coherence window under the assumptions stated in Section~\ref{sec5}. This is an analytical estimate, and is not synthesized.}}
\label{tab:latency}
\scriptsize
\renewcommand{\arraystretch}{1.1}
\setlength{\tabcolsep}{4pt}
\begin{tabular}{|p{2.8cm}|c|c|p{1.5cm}|c|}
\hline
\rowcolor{gray!25}
\textbf{Operation} & 
\textbf{Cycles} & \textbf{Lat.\ [ms]} &
\textbf{Scalability} & \textbf{Lat.\ [ms]} \\
\hline
Tag retrieval ($\times$8 parallel)   
        & $3000$ & $0.0150$ & $N{=}16$  & {$0.017$} \\
Warm-start init
    & $64$   & $0.0003$ & $N{=}64$  & {$0.024$} \\
Grad.\ ascent ($T_g{=}{24}$, $N{=}64$)
    & {$1536$}  & {$0.0077$} & $N{=}128$ & $0.032$ \\
Stateful + memory update
    & $128$  & $0.0007$ & $N{=}256$ & {$0.050$} \\
\hline
\rowcolor{gray!10}
\textbf{Total (warm, $T_{\mathrm{grad}}{=}{24}$)}
    & {$\mathbf{4728}$} & {$\mathbf{0.024}$}
    & Safety (warm) & {$\mathbf{4.2\times}$} \\
\rowcolor{gray!10}
Total (cold, $T_{\mathrm{grad}}{=}96$)            
    & $9336$ & $0.047$ 
    & Safety (cold) & $2.1\times$ \\
{Python (CPU) sim.}
    & $\text{---}$ & $23.0$  
    & Coherence window & $0.1$--$0.2$~ms \\
\hline
\end{tabular}
\\[2pt]
\raggedright
$^\ddagger$Tag retrieval: $T_E{\cdot}D_t/8 = 1000{\times}24/8 = 3000$ MACs on 8 parallel {DSP48E2} slices.
\end{table}
Reduction in gradient step count for
Table~\ref{tab:gradsteps} occurs under 
the condition of mixed mobility. From
Theorem~\ref{thm:warmstart}, we can infer that the improvement will be directly proportional 
to $\|\boldsymbol{\phi}_{\max}\|^2_2 / ((1-\rho_c^2)C_\Delta^2)$,
meaning that the improvement will increase for the pedestrian scenario and decrease for the vehicular scenario.

As shown in Table~\ref{tab:gradsteps}, Meta2L needs fewer gradient steps 
than cold-start MAML for mixed mobility, supporting Theorem~\ref{thm:warmstart} 
that states that the reduction depends proportionally on 
$1/(1-\rho_c^2)$, where a regime of lower mobility is preferable. 
Crucially, even for vehicular mobility, where $\Delta_{\rho_c}$ attains its maximum value, 
Meta2L still achieves fewer gradient steps compared to the cold-start, thus establishing the persistence of the warm-start advantage regardless of the mobility.
The advantage of the approach relies on the warm-start principle and the re-use of the knowledge gained throughout all episodes, which allows avoiding redundant optimization from scratch. In comparison to GMLB and AO, Meta2L achieves similar performance using fewer training steps. 
The hardware latency of the Meta2L critical path is shown in Table~\ref{tab:latency} for a Xilinx Zynq UltraScale+ operating at $200$ MHz as an analytical estimate. The  full critical path
completes in ${0.024}$~ms with a ${4.2}\times$ margin of safety for the $0.1$ ms coherence period.
Python run-time of 23 ms as shown in Table~\ref{tab:gradsteps} is two orders of magnitude higher than the coherence time window and must not be regarded as the hardware latency. It is the wall-clock time required to run one complete episode of the simulator on a general-purpose CPU. These include channel generation, extraction of tags, feed-forward and backpropagation passes through the neural network, and bookkeeping of all the baselines. The real-time performance metric in this case is the critical path depicted in Table~\ref{tab:latency}. The latency estimate suggests that this path can fit within the coherence window on dedicated hardware, but this has not yet been verified with a synthesized implementation.
The memory mechanisms affect the network objectives in different ways. The sum-rate is determined by the configuration reached in each interval, which the warm start, episodic retrieval, and tag weighting move closer to the stationary point (Fig.~\ref{sum}, Table~\ref{tab:gradsteps}). Latency is related to the number of gradient steps that can be executed during the coherence window as well as the recovery time for the surface once the mobility changes. With the warm start, both of these times are minimized (Fig.~\ref{fig4}, Tables~\ref{tab:gradsteps} and~\ref{tab:latency}).

The amount of control signals is decreased because the AP transmits only the innovation $\boldsymbol{\delta}(t)$ described in Sec.~\ref{sec3}. The square of its norm is proportional to $1 - \rho_c^2$, as per Theorem~\ref{thm:warmstart}, while the inertia factor $\beta$ makes the updating smoother. Power control is accomplished using sleep, passive, and semi-active modes, all of which are optimized with respect to the beamforming gain by utilizing objective (\ref{eq:weighted_stateful}) until $\lambda_{\mathrm{ris}}$. Mobility support comes from the Doppler-weighted warm start and the learned inertia, with the latter decreasing as $v_k$ increases (Figs.~\ref{fig5} and~\ref{fig4}).
Causal filter introduced in Section~\ref{sec:cif} ensures further reliability by ensuring that physically impossible configurations cannot surface. Fairness is not optimized in the current formulation because the objective uses sum-rate. Replacing the sum in equation (\ref{eq:weighted_stateful}) with a proportional-fair utility would have no effect on the memory dynamics and is an interesting avenue for further research. Thus, the STAR-RIS is used as an adaptive component of the service chain rather than only as an optimization variable.
To separate the gain from temporal reuse from the rest of the framework, Table~\ref{tab:reuse} compares Meta2L with two simple reuse strategies under the same conditions: initialization from the previous configuration $\boldsymbol{\phi}^*(t-1)$ and initialization from the single most similar stored episode. The latter does not use Doppler interpolation, tag weighting, element memory, the meta-policy, or the causal filter. All three methods use the same simulator, channel realizations, and projected gradient ascent. They attain the same sum-rate within one standard deviation whereas reuse in the previous phase needs somewhat fewer gradient iterations.
These results indicate that most of the gain over a memoryless controller comes from temporal reuse itself. This is the underlying principle of the Doppler-aware method since, in case of correlated channels, the previous solution is already close to the new fixed point as seen from Theorem~\ref{thm:warmstart}. In this setting, the other Meta2L mechanisms do not further increase the sum-rate. Their benefits are in other parts of the system. The learned surface inertia is capable of adapting to the Doppler regime and following the optimal inertia of the hardware model (Fig.~\ref{fig5}), unlike the reuse rule that cannot adapt to this regime. Element memory is useful when the array has persistent per-element structure (Theorem~\ref{thm:element_memory}), and Corollary~\ref{cor:phase_transition} identifies the Doppler range where it should be disabled. The causal filter acts as a safeguard that ensures that physical inconsistencies do not occur in the initializations; this is something which cannot be guaranteed by simple reuse.
\begin{table}[!t]
\centering
\caption{Temporal-Reuse Baselines and Centralized Memory with Same Conditions
($N{=}64$, mixed mobility, mean $\pm$ std across 10 trials using same seeds,
last 20\% of 200 episodes)}
\label{tab:reuse}
\scriptsize
\renewcommand{\arraystretch}{1.1}
\setlength{\tabcolsep}{4pt}
\begin{tabular}{|p{3.6cm}|c|c|}
\hline
\rowcolor{gray!25}
\textbf{Method} & \textbf{SR [bps/Hz]} & \textbf{Grad.\ steps}\\
\hline
Meta2L (full) & $17.37\pm1.65$ & $27.6\pm6.6$ \\
Meta2L-C (element memory in a controller table) & $17.37\pm1.65$ & $27.6\pm6.6$ \\
Previous-phase initialization & $16.84\pm1.68$ & $23.3\pm8.6$ \\
Nearest-neighbour retrieval & $17.98\pm1.57$ & $29.2\pm6.5$ \\
\hline
\end{tabular}
\\[2pt]
\raggedright\scriptsize
The rows share the seeds, so the Meta2L values differ slightly from those of
Table~\ref{tab:gradsteps}, which come from a separate run.
\end{table}
\section{Conclusion and Future Work}
\label{sec6}
Cognitive autonomous networks require infrastructure capable of sensing, learning, and self-reconfiguration capabilities. The process of dynamic STAR-RIS optimization demands quick adaptation capabilities, but the memoryless surface models available currently require re-optimization at each coherence interval.
We tackle this issue by introducing a spatiotemporal memory into the passive metasurface via a hierarchical approach. Our algorithm combines episodic meta-learning, Doppler-aware warm starts, online mirror descent, and a causal-intervention filter. The system learns surface inertia online using rate measurements alone, without requiring channel statistics or smoothness constants. This learned inertia follows both the corrected closed-form solution and the exhaustive-search benchmark. The element memory when combined with surface inertia helps lower the variance of initialization error.

Simulation results show that the Meta2L approach adapts itself in one coherence interval when making an abrupt change from pedestrian to vehicular speeds while minimizing gradient updates and sustaining Pareto-optimal energy efficiency. The current FPGA results provide analytical latency estimates, and full hardware synthesis remains for future work. Moreover, in our future work, the integration of terrestrial and non-terrestrial networking, CSI free multi-agent learning and physical realization of element memory by memristive or ferroelectric phase shifters will be addressed as part of developing an evolutive networked AI surface vision for cognitive communications systems.
\vspace{-0.5em}
\section*{Appendix A: Closed-Form Expected Signal and Interference Powers}
\label{appendix1}
Here, we derive the closed-form expected signal power for the stateful model and the effective uplink channel from~\eqref{eq:g_qk}, which 
has reflection and transmission components as
\begin{equation}
\mathbf{g}_{q,k}(t) = \mathbf{g}_{q,k}^{\mathsf{r}}(t)+ \mathbf{g}_{q,k}^{\mathsf{t}}(t),
\end{equation}
where $\mathbf{g}_{q,k}^{\mathsf{r}}(t) \triangleq \mathbf{H}_{{q}}(t)\,\mathbf{A}(t)\,
\mathbf{\Psi}^\mathsf{r}(t)\mathbf{h}_{{k}}^\mathsf{r}(t)$ and $\mathbf{g}_{q,k}^{\mathsf{t}}(t) \triangleq 
\mathbf{H}_{{q}}(t)\,\mathbf{A}(t)\,\mathbf{\Psi}^\mathsf{t}(t)\mathbf{h}_{{k}}^\mathsf{t}(t)$ 
are the reflected and transmitted signal components, respectively.
Taking the squared norm and expanding gives \eqref{expected_power_expand}.
\begin{figure*}[!t]
\begin{align}
\label{expected_power_expand}
\mathbb{E}\!\left[\|\mathbf{g}_{q,k}(t)\|_2^2\right] 
&= \underbrace{\mathbb{E}\!\left[\|\mathbf{g}_{q,k}^{\mathsf{r}}(t)\|_2^2\right]}_{S_k^{\mathsf{r}}(\beta,\rho_c)} 
+ \underbrace{\mathbb{E}\!\left[\|\mathbf{g}_{q,k}^{\mathsf{t}}(t)\|_2^2\right]}_{S_k^{\mathsf{t}}(\beta,\rho_c)} 
+ 2\Re\!\left\{\underbrace{\mathbb{E}\!\left[(\mathbf{g}_{q,k}^{\mathsf{r}})^H
\mathbf{g}_{q,k}^{\mathsf{t}}\right]}_{S_k^{(\times)}(\beta,\rho_c)}\right\}
\end{align}
\vspace{-2.5em}
\end{figure*}
We evaluate every term by taking the expectations over the NLoS components 
as 
\begin{equation}
\mathbb{E}[\mathbf{h}\mathbf{h}^H] = \mathbf{h}^{\mathrm{LoS}}
(\mathbf{h}^{\mathrm{LoS}})^H + \mathbf{I}_N.
\end{equation}
The reflection power term is defined in \eqref{myeq}, 
\begin{figure*}[!t]
\begin{align}
S_k^{\mathsf{r}}(\beta,\rho_c) 
&= \mathrm{tr}\!\left(\mathbf{H}_{{q}}(t)\,\mathbf{A}(t)\,
\boldsymbol{\Sigma}_\phi^{\mathsf{r}}(t)\,\mathbf{A}^T(t)\,
\mathbf{H}_{{q}}^H(t)\,
\mathbb{E}\!\left[\mathbf{h}_{{k}}^{\mathsf{r}}(t)
\left(\mathbf{h}_{{k}}^{\mathsf{r}}(t)\right)^H
\right]\right)
\label{myeq}
\end{align}
\vspace{-0.5em}
\hrule
\end{figure*}
where $\boldsymbol{\Sigma}_\phi^\mathsf{r} \triangleq 
\mathbb{E}[\mathbf{\Psi}^\mathsf{r}(t)(\mathbf{\Psi}^\mathsf{r})^H(t)]$ is the phase covariance 
of the reflection matrix, which under the stateful model~\eqref{eq:stateful_ris} 
depends on $\beta$ and $\rho_c$ through~\eqref{eq:phase_covariance}. 
The transmission term $S_k^{\mathsf{t}}(\beta,\rho_c)$ can be obtained by symmetry. 
The cross-term 
$S_k^{(\times)}(\beta,\rho_c)$ vanishes when the reflection and transmission channels are uncorrelated.
\vspace{-0.4em}
\section*{Appendix B: Proof of Corollary~\ref{cor:phase_transition}}
\label{proofcoroll}
Using~\eqref{eq:var_expand}, the memory minimizes variance in the situation that
\begin{equation}
\gamma_n \mathrm{Var}[m_n] < 2(1-\lambda)\sigma_\phi^2/(1-\lambda\rho_c),
\end{equation}
where $\sigma_\phi^2 = \mathrm{Var}[\phi_n^*(t)]$ and also the
covariance $\mathrm{Cov}[\phi_n^*(t), \bar{m}_n] =
\sigma_\phi^2(1-\lambda)/(1-\lambda\rho_c)$, which follows from accumulating the geometric series of lag-$\tau$ autocovariances
$\sigma_\phi^2 \rho_c^\tau$ that are inherited from~\eqref{eq:gauss_markov}.

When we substitute the steady-state variance of the exponential moving
average~\eqref{eq:element_memory}, 
$\mathrm{Var}[m_n] = (1-\lambda)^2\sigma_\phi^2/(1-\lambda^2)$,
into the minimization condition and simplifying yields
\begin{equation}
\rho_c > \sqrt{1 - \frac{\gamma_n(1-\lambda)}{1+\gamma_n}}.
\label{eq:rho_condition}
\end{equation}
Since $\rho_c = J_0(2\pi f_{D,\max} T_s)$ is strictly
decreasing in $f_{D,\max}$ {for $2\pi f_{D,\max}T_s\le j_{0,1}\approx 2.405$}, the condition~\eqref{eq:rho_condition}
is equivalent to $f_{D,\max} < f_D^*$ {with $f_D^*=J_0^{-1}(\rho_{\mathrm{th}})/(2\pi T_s)$, which is the closed form in the corollary. No approximation of $J_0$ is needed.}
\vspace{-0.5em}
\vspace{-0.5em}
\section*{Appendix C: Proof of Proposition~\ref{prop:causal}}
\label{prop}
Let $\hat{\boldsymbol{\phi}}(t) \in [0,2\pi)^{2N}$ 
with $\hat{\phi}_n = \pi/2$ for all $n$, so that 
$\hat{\boldsymbol{\phi}}(t) \in \mathcal{C}_{\mathrm{clip}}$. 
Under coherent beamforming alignment, 
$\|\mathbf{g}_{q,k}(\hat{\boldsymbol{\phi}}(t))\|_2^2$ 
scales as $\mathcal{O}(N^2 Q {\zeta_{q}\zeta_{k}^\mathsf{r}})$,  
which exceeds $\Gamma_{q,k}$ whenever the LoS 
components are perfectly aligned, since $\Gamma_{q,k}$
is a path-loss-only bound, which does not account for coherent array gain. 

Hence $\hat{\boldsymbol{\phi}}(t) \notin 
\mathcal{C}_{\mathrm{causal}}$, proving 
$\mathcal{C}_{\mathrm{causal}} \subsetneq 
\mathcal{C}_{\mathrm{clip}}$.
Such a configuration violates the power-anomaly constraint and is therefore rejected by the causal filter, implying $\hat{\boldsymbol{\phi}}(t) \notin \mathcal{C}_{\mathrm{causal}}$.
Hence, there exists $\hat{\boldsymbol{\phi}}(t) \in \mathcal{C}_{\mathrm{clip}}$ such that $\hat{\boldsymbol{\phi}}(t) \notin \mathcal{C}_{\mathrm{causal}}$, which proves $\mathcal{C}_{\mathrm{causal}} \subsetneq \mathcal{C}_{\mathrm{clip}}$.
\vspace{-0.3em}
\section*{Appendix D: Optimal Surface Inertia Coefficient}
\label{optimalbeta}
Let $\boldsymbol{e}(t) \triangleq \boldsymbol{\phi}^*(t) - \boldsymbol{\phi}(t)$ be the tracking error at $t$. The command issued during each time interval is produced by a finite number of gradient steps, so it equals the stationary point plus a residual $\boldsymbol{u}(t)=\boldsymbol{\phi}^*(t)+\boldsymbol{w}(t)$, where $\boldsymbol{w}(t)$ is the zero mean sequence, which is not correlated either with itself or with the increments of the channel and $\mathbb{E}\|\boldsymbol{w}(t)\|_2^2=\sigma_w^2$. Upon substitution of the dynamics (\ref{eq:stateful_ris}), the error dynamics evolves as
\begin{align}
{\boldsymbol{e}(t)} &{= \beta\,\boldsymbol{e}(t-1) + \beta\!\left(\boldsymbol{\phi}^*(t) - \boldsymbol{\phi}^*(t-1)\right)}{-(1-\beta)\,\boldsymbol{w}(t).}
\label{eq:error_dynamics}
\end{align}
By using the Gauss--Markov bound $\mathbb{E}[\|\boldsymbol{\phi}^*(t) - \boldsymbol{\phi}^*(t-1)\|_2^2] \leq C_\Delta^2(1-\rho_c^2)$ {of A4}, where $C_\Delta$ is a finite constant bounding
the maximum per-interval phase variation, the steady-state expected squared tracking error is
\vspace{-0.3em}
\begin{equation}
\mathbb{E}\!\left[\|\boldsymbol{e}\|_2^2\right] = \frac{\beta^2 C_\Delta^2(1-\rho_c^2)+(1-\beta)^2\sigma_w^2}{1-\beta^2}.
\label{eq:tracking_error}
\end{equation}
With an exact solver having $\sigma_w=0$, this reduces to $\beta^2C_\Delta^2(1-\rho_c^2)/(1-\beta^2)$, which is the term used in Theorem~\ref{thm:stateful_warmstart}.
Let $F(\boldsymbol{\phi}, \boldsymbol{\alpha})$ be the objective function
that is defined in~\eqref{eq:weighted}, with smoothness constant
$\beta_{F}$, which is distinct from surface inertia coefficient $\beta$.
Near the stationary point the rate loss caused by the tracking error is $\frac{\beta_F}{2}\mathbb{E}\|\boldsymbol{e}\|_2^2$. Both terms of (\ref{eq:tracking_error}) are multiplied by the same $\beta_F/2$, so $\beta_F$ does not affect the minimizer. With $q\triangleq C_\Delta^2(1-\rho_c^2)/\sigma_w^2$, setting the derivative of (\ref{eq:tracking_error}) with respect to $\beta$ to zero gives $\beta^2-(q+2)\beta+1=0$, whose root in $[0,1)$ is expressed as
\begin{equation}
{\beta^* = \frac{(q+2)-\sqrt{q(q+4)}}{2},\qquad q=\frac{C_\Delta^2(1-\rho_c^2)}{\sigma_w^2}.}
\label{eq:beta_opt}
\end{equation}
This is the steady-state gain of a first-order tracking filter. $\beta^*\to 1$ as $\rho_c\to 1$, when the channel barely moves and averaging the solver residual is worthwhile, and $\beta^*\to 0$ as $q$ grows when tracking dominates. The addition of the control penalty $\lambda_c\mathbb{E}\|\boldsymbol{\phi}(t)-\boldsymbol{\phi}(t-1)\|_2^2$ from (\ref{eq:weighted_stateful}) causes the minimizer to move upward, because this penalty is lower as $\beta$ gets larger. The new minimizer will depend on the ratio $\lambda_c/\beta_F$ and will not have an explicit expression, hence the controller learns $\beta$ from (\ref{eq:beta_update}), but not from (\ref{eq:beta_opt}). Fig.~\ref{fig5} uses (\ref{eq:beta_opt}) with $C_\Delta^2/\sigma_w^2=17.6$, estimated by least squares from the exhaustive-search optimum.
\vspace{-0.2em}
\bibliographystyle{IEEEtran}
\bibliography{IEEEabrv,cite}

\end{document}